\documentclass[sigconf]{acmart}
\AtBeginDocument{%
  }

\setcopyright{acmlicensed}
\copyrightyear{2025}
\acmYear{2025}
\acmDOI{XXXXXXX.XXXXXXX}
\acmConference[AsiaCCS '26]{21st ACM ASIA Conference on Computer and Communications Security}{June 01--06,
  2025}{Bangalore, India}
\acmISBN{978-1-4503-XXXX-X/2018/06}

\usepackage{algorithmic}
\usepackage{graphicx}
\usepackage{textcomp}
\usepackage{xcolor}
\usepackage[table]{xcolor}
\usepackage{enumitem}
\setlist[enumerate]{itemsep=0mm}
\usepackage{hyperref}
\usepackage{multirow}
\usepackage{pifont}
\usepackage{framed}
\usepackage{graphicx}    
\usepackage{subcaption}    
\usepackage{caption}       
\usepackage[justification=centering]{caption}
\usepackage{threeparttable}
\usepackage{booktabs}

\usepackage[normalem]{ulem}
\useunder{\uline}{\ul}{}
\usepackage{tabularx}
\usepackage{adjustbox}
\usepackage{caption}
\usepackage{makecell}
\usepackage{tikz}
\usepackage[most]{tcolorbox}
\definecolor{true}{RGB}{91, 181, 172}  
\definecolor{false}{RGB}{222, 82, 108}
\newcommand{\blackcheck}{\ding{51}}
\newcommand{\blackcross}{\ding{55}}
\newcommand{\greencheck}{\textcolor{true}{\blackcheck}}         
\newcommand{\redcross}{\textcolor{false}{\blackcross}}                 
\newcommand{\yellowcircle}{%
  \tikz \draw[line width=1pt, color=yellow!50!orange] (0,0) circle (0.1cm);%
}

\begin{document}

\title{Revisiting Pre-trained Language Models for Vulnerability Detection}

\author{Youpeng Li}
\affiliation{%
  \institution{University of Texas at Dallas}
  \city{Richardson}
  \state{Texas}
  \country{USA}
}
\email{youpeng.li@utdallas.edu}

\author{Weiliang Qi}
\affiliation{%
  \institution{University of Texas at Dallas}
  \city{Richardson}
  \state{Texas}
  \country{USA}
}
\email{weiliang.qi@utdallas.edu}

\author{Xuyu Wang}
\affiliation{%
  \institution{Florida International University}
  \city{Miami}
  \state{Florida}
  \country{USA}
}
\email{xuywang@fiu.edu}

\author{Fuxun Yu}
\affiliation{%
  \institution{Microsoft}
  \city{Redmond}
  \state{Washington}
  \country{USA}
}
\email{fuxunyu@microsoft.com}

\author{Xinda Wang}
\affiliation{%
  \institution{University of Texas at Dallas}
  \city{Richardson}
  \state{Texas}
  \country{USA}
}
\email{xinda.wang@utdallas.edu}

\renewcommand{\shortauthors}{Li et al.}

\begin{abstract}
The rapid advancement of pre-trained language models (PLMs) has demonstrated promising results for various code-related tasks. However, their effectiveness in detecting real-world vulnerabilities remains a critical challenge. While existing empirical studies evaluate PLMs for vulnerability detection (VD), they suffer from data leakage, limited scope, and superficial analysis, hindering the accuracy and comprehensiveness of evaluations. This paper begins by revisiting the common issues in existing research on PLMs for VD through the evaluation pipeline. It then proceeds with an accurate and extensive evaluation of 18 PLMs, spanning model parameters from millions to billions, on high-quality datasets that feature accurate labeling, diverse vulnerability types, and various projects. Specifically, we compare the performance of PLMs under both fine-tuning and prompt engineering, assess their effectiveness and generalizability across various training and testing settings, and analyze their robustness to perturbations such as code normalization, abstraction, and semantic-preserving transformations.

Our findings reveal that, for function-level VD, PLMs incorporating pre-training tasks designed to capture the syntactic and semantic patterns of code outperform both general-purpose PLMs and those solely pre-trained or fine-tuned on large code corpora. However, these models face notable challenges in real-world scenarios, such as difficulties in detecting vulnerabilities with complex dependencies, handling perturbations introduced by code normalization and abstraction, and identifying semantic-preserving vulnerable code transformations. Also, the truncation caused by the limited context windows of PLMs can lead to a non-negligible number of labeling errors, which is overlooked by previous work. This study underscores the importance of thorough evaluations of model performance in practical scenarios and outlines future directions to help enhance the effectiveness of PLMs for realistic VD applications.
\end{abstract}




\maketitle

\section{Introduction}
Pre-trained language models (PLMs) are transforming software development by helping developers automate repetitive tasks such as code completion and summarization. However, enhancing the ability of PLMs to detect complex, diverse, and subtle real-world vulnerabilities remains a critical challenge.

Although existing research has explored if PLMs can demonstrate strong performance in vulnerability detection (VD), limitations across various stages of the evaluation pipeline hinders an accurate reflection of PLMs’ true capabilities: 
\textit{(i) Data Leakage}. Most studies rely on evaluation datasets that inherently introduce data leakage, leading to biased estimations of the model performance in real-world scenarios. The randomly partitioning method used in existing studies~\cite{diversevul,a16,a17,a18,a19,a20,benjamin2023,a22,ni1,yinxin,a28,a29,davidlo,a31,a32} often causes models to encounter duplicated code patterns between the training and test data.
The temporal overlap between the pre-training data cut-off dates of PLMs and the commit dates of evaluation data is also frequently overlooked~\cite{a26,a27,a28,a29,davidlo,a31,a32}.
\textit{(ii) Limited Scope}. The experimental setup and settings adopted by many studies are neither comprehensive nor aligned with real-world scenarios, resulting in unrepresentative conclusions. For example, some studies focus on models with constrained architectures and parameter scales~\cite{graph_prompt,diversevul,a16,a17}. Also, models are often evaluated on samples with limited size, a narrow range of vulnerability types~\cite{benjamin,ullah,a10,a11,a12,a13,a28,a29,davidlo}, or an unrealistically balanced distribution~\cite{a6,a7,a8,a9,a10,a11,a12,a13}. Others focus exclusively on either fine-tuning~\cite{diversevul,a16,a17,a18,a19,a20,benjamin2023,a22} or prompting engineering~\cite{a6,a7,a8,a9,a10,a11,a12,a13}, neglecting comparative analyses across different adaptation methods or considerations of the trade-offs between cost and performance of PLMs in VD. 
\textit{(iii) Superficial Analysis}. Existing studies~\cite{yinxin,a26,a27,a28,a29,davidlo,a31,a32} primarily focus on performance comparisons, without thoroughly investigating how practical factors (e.g., code normalization, abstraction, transformation, truncation) influence the effectiveness of PLMs for VD. This creates a significant gap between performance estimates and real-world applications, failing to provide insights for enhancing PLMs' true capabilities.

Therefore, this paper \underline{Revisit}s the capabilities of PLMs for \underline{VD} (\textbf{RevisitVD}) through an extensive and realistic evaluation that addresses the shortcomings of existing evaluation works. In particular, starting from data preparation, we discuss the limitations of existing work in selecting evaluation datasets, including insufficient consideration of data volume, the diversity of vulnerability types, and inherent labeling errors within the datasets. To address these issues, we introduce our reconstructed dataset, alongside a time-order-based dataset partitioning method to avoid data leakage caused by the random data partitioning commonly used in current VD research. Additionally, we highlight the risk that existing VD datasets may predate the cutoff date of PLMs’ pre-training data, potentially leading to data leakage. To mitigate this, we collect a new C/C++ function-based VD dataset from NVD~\cite{nvd}, encompassing various vulnerability types and projects, with all samples having commit dates after the pre-training cutoff dates of PLMs evaluated.

To ensure the comprehensiveness and representativeness of the evaluation, we evaluate 18 PLMs with parameter sizes ranging from millions to billions, covering a variety of model architectures. All models have been specifically pre-trained on code structure-aware tasks or exposed to large-scale code corpora, making their evaluation on VD tasks particularly relevant and competitive. Standing out from other existing evaluation efforts, we comprehensively compare PLMs' performance using two model adaptation techniques: fine-tuning and prompt engineering. 
Specifically, we fully fine-tune small language models (i.e., BERT series and CodeT5), while applying LoRA for partial fine-tuning of LLMs with up to 34B parameters. For prompting engineering, we adopt both zero-shot and few-shot prompting. Within each prompt setting, in addition to raw code functions, we introduce three types of structural and semantic-aware prompts: flattened abstract syntax tree, code with API calls, and code with data flow. These prompts embed structural information and dependencies within the code, guiding PLMs to better analyze vulnerabilities. We also introduce the existing state-of-the-art (SOTA) reasoning model, GPT-5 Mini, to explore the improvements that Chain-of-Thoughts (CoT) brings to VD.
Furthermore, we evaluate the performance of PLMs on out-of-distribution data and test data under various perturbations (i.e., normalization, abstraction, semantic-preserving transformations) to examine their generalizability and robustness in practical applications.

Through experimental analysis, RevisitVD reveals the following findings: (1) PLMs pre-trained on specialized tasks that guide them in learning code syntax and semantics (e.g., PDBERT) significantly outperform most PLMs pre-trained or fine-tuned on general language modeling (e.g., CodeLlama), despite having far fewer parameters. This suggests that future research could focus on designing specialized pre-training tasks to improve PLM capabilities for VD while considering the trade-off between performance and efficiency. (2) Evaluating fine-tuned PLMs on test data from the same source as the training data may lead to inaccurate assessments of their real-world capabilities. (3) Existing PLMs struggle with detecting vulnerabilities that involve complex program dependencies, and lack robustness to minor perturbations, such as inconsistencies in normalization rules applied during training and testing. (4) Most PLMs demonstrate certain robustness to abstracted code, suggesting that their predictions do not rely solely on textual words. (5) Most PLMs exhibit varing degrees of performance drop in semantic-preserving transformations, indicating that they are not yet reliable against vulnerable code reuse or adversarial examples. (6) The limited context window size of PLMs unintentionally introduces label errors during truncation, disrupting model training. Code slicing that reduces input length can help PLMs focus more effectively on learning vulnerability patterns.

In summary, we make the following contributions:

(1) We revisit the true capabilities of PLMs for VD by addressing multiple critical shortcomings of existing research. These include data leakage caused by flawed dataset partitioning and overlooking the temporal overlap between evaluation data and pre-training knowledge of LLMs; limited scope resulting from the evaluation of constrained evaluation setups on unrepresentative experimental settings; and superficial analysis neglecting factors like model generalization and robustness to practical use.

(2) We evaluate a wide range of 18 PLMs, from small to large parameter scales, and compares their performance using two adaptation techniques: fine-tuning and prompt engineering. These comprehensive comparisons, which also includes the use of structural and semantic-aware prompts and the CoT reasoning model, offer a realistic estimation of PLMs' capabilities for VD.

(3) We implement an extensible evaluation framework built on newly collected, high-quality datasets that go beyond LLMs' pre-training knowledge cutoff, allowing for future evaluation of stronger models on more recent data. We reveal several key insights into the capabilities and weaknesses of PLMs, along with effective strategies for enhancing their performance in practical settings. Our artifacts are available https://github.com/youpengl/RevisitVD.

    

\section{Background and Related Work}
\label{sec:section2}
\subsection{Pre-trained Language Models}
The Transformer architecture~\cite{attention} has become the foundation of most pre-trained language models, using self-attention to capture dependencies across entire sequences. It consists of an encoder and a decoder, which can be used independently or together depending on the task.

{\noindent{\textbf{Encoder-based Pre-trained Language Models.}}}
For encoder-based PLMs, one of the core pre-training tasks is masked language modeling~\cite{bert} where the model improves contextual understanding by learning to predict randomly masked tokens in the sequence. Given a token sequence $\mathcal{U}^{mask}=\{u_0, u_1^{mask}, \cdots, u_i^{mask}, \cdots, u_{m-1}\}$ with a certain percentage of randomly masked tokens, the optimization of the PLM $\Theta$ can be formulated as follows:
\begin{equation}
\Theta = \arg\max_\Theta \sum_{u_i\in S_{mask}}\log P(u_i \mid \mathcal{U}^{mask}; \Theta),
\end{equation}
\noindent where $S_{mask}$ is the set of masked tokens and $\Theta$ represents the parameters of the PLM, optimized through stochastic gradient descent (SGD).

{\noindent{\textbf{Decoder-based Pre-trained Language Models.}}}
For decoder-based PLMs, the primary objective during pre-training is causal language modeling~\cite{GPT} where the model improves its understanding of language patterns by learning to predict the next token based on preceding context. Given a sequence of tokens $\mathcal{U}=\{u_0, u_1, \cdots, u_i, \cdots, u_{m-1}\}$, the optimization of the PLM $\Theta$ can be expressed as follows:

\begin{equation}
\Theta = \arg \max_\Theta \sum_i \log P(u_i \mid u_{i-n}, \ldots, u_{i-1}; \Theta),
\end{equation}

where $n$ denotes the size of the context window, and $\Theta$ represents the parameters of the PLM.

\begin{table*}[h]

\caption{Comparison of vulnerability detection benchmarks}
\label{tab:literature}
  \centering
\resizebox{0.75\textwidth}{!}{
\begin{threeparttable}
\begin{tabular}{cccccccccc}
\toprule
& \textbf{\makecell{Model\\Diversity}} & \textbf{\makecell{Data\\Diversity}} & \textbf{\makecell{Time\\Split}} & \textbf{\makecell{Balanced\\Training}} & \textbf{\makecell{Realistic\\Testing}} & \textbf{\makecell{Knowledge\\Cutoff}} & \textbf{\makecell{Finetuned\\Model Size}} & \textbf{\makecell{Out of \\Distribution}} & \textbf{\makecell{Robust\\Analysis}} \\
\midrule
\multicolumn{10}{l}{\textbf{Prompt Engineering-based Evaluation}} \\
\midrule
Khare et al~\cite{understanding}  & \greencheck & \greencheck & N/A & N/A & \redcross & \redcross & N/A & N/A & N/A\\
Steenhoek et al~\cite{benjamin}  & \greencheck & \redcross & N/A & N/A & \redcross & \redcross & N/A & N/A & N/A\\
SecLLMHolmes~\cite{ullah}  & \greencheck & \redcross & N/A & N/A & \redcross & \greencheck & N/A & N/A & N/A\\
CORRECT~\cite{a6}  & \greencheck & \redcross & N/A & N/A & \redcross & \redcross & N/A & N/A & N/A\\
VulnSage~\cite{a7}  & \greencheck & \greencheck & N/A & N/A & \redcross & \greencheck & N/A & N/A & N/A\\
VulnLLMEval~\cite{a8}  & \greencheck & \redcross & N/A & N/A & \redcross & \greencheck & N/A & N/A & N/A\\
VulDetectBench~\cite{a9}  & \greencheck & \greencheck & N/A & N/A & \redcross & \redcross & N/A & N/A & N/A\\
VulBench~\cite{a10}  & \greencheck & \redcross & N/A & N/A & \redcross & \redcross & N/A & N/A & N/A\\
Steenhoek et al~\cite{a11}  & \greencheck & \redcross & N/A & N/A & \redcross & \redcross & N/A & N/A & N/A\\
LLM4Vuln~\cite{a12}  & \greencheck & \redcross & N/A & N/A & \redcross & \greencheck & N/A & N/A & N/A\\
SecureFalcon~\cite{a13}  & \greencheck & \redcross & N/A & N/A & \redcross & \redcross & N/A & N/A & N/A\\
Zhang et al~\cite{graph_prompt}  & \redcross & \greencheck & N/A & N/A & \greencheck & \redcross & N/A & N/A & N/A\\
\midrule
\multicolumn{8}{l}{\textbf{Fine Tuning-based Evaluation}} \\
\midrule
DiverseVul~\cite{diversevul} & \redcross & \greencheck & \redcross& \redcross & \greencheck & \redcross & $\sim$ 220M & \greencheck & \redcross \\
Thapa et al~\cite{a16} & \redcross & \redcross & \redcross& \greencheck & \redcross & \redcross & $\sim$ 6B & \redcross & \redcross \\
CleanVul~\cite{a17} & \redcross & \greencheck & \redcross& \greencheck & \redcross & \redcross & $\sim$ 7B & \greencheck & \redcross \\
VulLLM~\cite{a18} & \greencheck & \greencheck & \redcross& \greencheck & \redcross & \redcross & $\sim$ 15B & \greencheck & \greencheck \\
VulnPatchPairs~\cite{a19} & \redcross & \redcross & \redcross& \yellowcircle & \greencheck & \redcross & $\sim$ 220M & \greencheck & \greencheck \\
Aleksei et al~\cite{a20} & \redcross & \greencheck & \redcross& \yellowcircle & \greencheck & \redcross & $\sim$ 13B & \redcross & \redcross \\
Steenhoek et al~\cite{benjamin2023} & \redcross & \redcross & \redcross& \yellowcircle & \greencheck & \redcross & $\sim$ 125M & \greencheck & \redcross \\
Jiang et al~\cite{a22} & \greencheck & \greencheck & \redcross& \yellowcircle & \greencheck & \redcross & $\sim$ 8B & \redcross & \redcross \\
\midrule
\multicolumn{8}{l}{\textbf{Considering Both Prompt Engineering- and Fine Tuning-based Evaluation}} \\
\midrule
Ni et al~\cite{ni1} & \redcross & \greencheck & \redcross& \redcross & \greencheck & \redcross & $\sim$ 125M & \redcross & \greencheck \\
PrimeVul~\cite{primevul} & \greencheck & \greencheck & \greencheck& \redcross & \greencheck & \redcross & $\sim$ 175B & \greencheck & \redcross \\
Yin et al~\cite{yinxin} & \greencheck & \greencheck & \redcross& \redcross & \greencheck & \redcross & $\sim$ 7B & \redcross & \redcross \\
Zhang et al~\cite{a26} & \greencheck & \greencheck & \greencheck& \redcross & \greencheck & \redcross & $\sim$ 8.5B & \redcross & \redcross \\
VulEval~\cite{a27} & \greencheck & \greencheck & \greencheck & \redcross & \redcross & \redcross & $\sim$ 220M & \redcross & \redcross \\
Purba et al~\cite{a28} & \redcross & \redcross & \redcross& \greencheck & \redcross & \redcross & $\sim$ 13B & \redcross & \redcross \\
Zhou et al~\cite{a29} & \greencheck & \redcross & \redcross& \greencheck & \greencheck & \redcross & $\sim$ 8B & \redcross & \redcross \\
Zhou et al~\cite{davidlo} & \redcross & \redcross & \redcross& \greencheck & \redcross & \redcross & $\sim$ 125M & \redcross & \redcross \\
ChatGPT4Vul~\cite{a31} & \redcross & \greencheck & \redcross& \yellowcircle & \greencheck & \redcross & $\sim$ 125M & \redcross & \redcross \\
Guo et al~\cite{a32} & \greencheck & \greencheck & \redcross& \yellowcircle & \greencheck & \redcross & $\sim$ 7B & \greencheck & \redcross \\
\textbf{RevisitVD (Ours)} & \greencheck & \greencheck & \greencheck& \yellowcircle & \greencheck & \greencheck & $\sim$ 34B & \greencheck & \greencheck \\
\bottomrule
\end{tabular}
\begin{tablenotes}[flushleft] 
\item \yellowcircle: considering both imbalanced and balanced training stettings
\end{tablenotes}
\end{threeparttable}
}
\vspace{-0.1in}
\end{table*}

\subsection{Pre-trained Language Models for Code}
Given the unique syntactic and structural characteristics of code, PLMs originally designed for natural language processing (NLP) tasks often require specialized adaptations when applied to code-related tasks. These adaptations include: (1) introducing specifically designed code structure-aware pre-training tasks to enhance code understanding, and (2) pre-training PLMs with the general language modeling objective on large code corpora that encompass source code from multiple programming languages. The former approach is typically implemented using encoder-based model architectures, with parameter sizes generally in the millions, while the latter often utilizes decoder-based model architectures, with parameter sizes reaching billions. In order to distinguish them, this paper refers to code-specific small PLMs with millions of parameters as \textbf{Code SLMs} and those with billions of parameters as LLMs. Additionally, LLMs can be further categorized as \textbf{Code LLMs} or general-purpose LLMs, depending on their intended application.

{\noindent{\textbf{Small Language Models for Code (Code SLMs).}}}
Code SLMs often integrate specialized, code-specific knowledge into their pre-training tasks. 
For instance, UniXCoder~\cite{UniXcoder} transforms the Abstract Syntax Tree (AST) of a function into a flattened AST using its proposed mapping algorithm, aiding the model in learning syntactic information within code. GraphCodeBERT~\cite{graphcodebert} integrates a data def-use prediction task during pre-training, enhancing the model’s capability to capture data flow. PDBERT~\cite{pdbert} introduces two pre-training tasks: statement-level control dependency prediction and token-level data dependency prediction, to guide the model in learning semantic relationships in code. CodeT5~\cite{codet5} involves an identifier-aware pre-training task to differentiate and recover identifiers for improved code understanding and generation.

{\noindent{\textbf{Large Language Models for Code (Code LLMs).}}}
For Code LLMs, the fill-in-the-middle approach~\cite{FIIM} is adopted by models like CodeLlama~\cite{codellama} and DeepSeek-Coder~\cite{deepseekcoder} to complement left-to-right generative capabilities of decoder-based PLMs, enhancing code generation and infilling. To handle long-context tasks such as cross-file code completion, many Code LLMs extend input sequence length by reconfiguring parameters in rotary position embeddings (RoPE)~\cite{ROPE}. To further improve Code LLMs’ ability for code-related tasks, numerous curated instruction datasets are used to fine-tune Code LLMs~\cite{starchat, wizardcoder}.

\vspace{-0.1in}
\subsection{Evaluating PLMs for Vulnerability Detection}

{\noindent{\textbf{Limitations in Existing Evaluations.}}} 
Existing works fall short in comprehensive and accurate evaluations, which limits our understanding of LLMs' applicability in real-world VD applications. These limitations stem from a limited scope, resulting from evaluations of constrained evaluation setups on unrepresentative experimental settings. They also arise from flawed data partitioning methods, overlooked temporal overlap between evaluation data and LLM pre-training knowledge, and a lack of analysis regarding model generalization and robustness in real-world scenarios.

First, many prior studies focus exclusively on either fine-tuning~\cite{diversevul,a16,a17,a18,a19,a20,benjamin2023,a22} or prompt engineering~\cite{understanding,benjamin,ullah,a6,a7,a8,a9,a10,a11,a12,a13,graph_prompt}, without conducting a comprehensive comparison or analyzing the trade-offs between performance and efficiency. Also, some evaluations suffer from biased evaluation results due to limitations in model size~\cite{graph_prompt,diversevul,a16,a17,a19,a20,benjamin2023,ni1,a28,davidlo,a31}, dataset scale, and vulnerability diversity~\cite{benjamin,ullah,a6,a8,a10,a11,a12,a13,a16,a19,benjamin2023,a28,a29,davidlo}. Furthermore, inappropriate data partitioning can skew training outcomes and produce unrealistic performance estimates. In real-world settings, VD datasets are often highly imbalanced. Fine-tuning under such conditions~\cite{diversevul,ni1,primevul,yinxin,graph_prompt} can bias models toward non-vulnerable samples, leading to high false negative rates. Conversely, evaluating models under artificially balanced conditions~\cite{understanding,benjamin,ullah,a6,a7,a8,a9,a10,a11,a12,a13,a16,a17,a18,a27,a28,davidlo} risks grossly overestimating precision, as false positives are just as critical in security analysis as false negatives. Another concern is the prevalent use of random data partitioning~\cite{diversevul,a16,a17,a18,a19,a20,benjamin2023,a22,ni1,yinxin,a28,a29,davidlo,a31,a32}, which risks data leakage if similar vulnerability patterns appear in both training and test sets. Potential temporal overlap ~\cite{understanding,benjamin,a6,a9,a10,a11,a13,graph_prompt,diversevul,a16,a17,a18,a19,a20,benjamin2023,a22,ni1,primevul,yinxin,a26,a27,a28,a29,davidlo,a31,a32} between LLMs' pretraining data cutoff dates and test data's commit dates can further exacerbate this issue. Moreover, evaluations conducted solely on in-distribution data~\cite{a16,a20,a22,ni1,yinxin,a26,a27,a28,a29,davidlo,a31} may overestimate model performance. Validating models on test data from diverse sources is essential to assess generalization. Finally, robustness analysis is equally important, as it evaluates how models withstand real-world adversarial challenges, such as detecting vulnerable code reuse under various perturbations.

{\noindent{\textbf{Our Work.}}} 
In contrast to limitations of existing evaluations as detailed in Table~\ref{tab:literature}, our study comprehensively addresses all of aforementioned issues at every stage through the evaluation pipeline. For the first time, we conduct the in-depth and accurate evaluation of 18 representative PLMs for VD, encompassing different scales of model parameters and systematically compares fine-tuning and prompt engineering on the high-quality datasets. For fine-tuning, we include open-source PLMs with up to 34 billion parameters. For prompt engineering, we evaluate two prompt settings and four types of structure- and semantic-aware prompts to determine better configurations. Finally, we assess the real-world performance of PLMs from multiple realistic perspectives, including their generalization to out-of-distribution data and robustness to code perturbations (normalization, abstraction, semantic-preserving transformations, truncation). Our analysis offers valuable insights to guide future research in VD using PLMs.
\section{Experimental Setup}
\subsection{Reconstructed Dataset}\label{reconstructed}
\subsubsection{Dataset Selection}

Despite efforts to build numerous VD datasets and benchmarks~\cite{bigvul,crossvul,diversevul,Devign, ReVeal}, limitations like limited data diversity/volume, unrealistic synthetic samples, balanced evaluation data distributions, and inaccurate labeling still persist, leading to biased evaluations in VD research.
Many existing datasets cover only a few CWE types or projects~\cite{VulDeePecker, Devign, ReVeal, DeepWukong}, or include synthetic data that poorly reflects real-world vulnerabilities~\cite{uVulDeePecker, juliet:online, SySeVR, VulDeeLocator}. While some datasets like SVEN~\cite{SVEN} and SecLLMHolmes~\cite{ullah} are collected from real-world projects and manually labeled, they are limited in scale and exhibit unrealistic class balance. More comprehensive datasets like Big-Vul\cite{bigvul}, CVEFixes~\cite{cvefixes}, PatchDB~\cite{wang2021patchdb}, CrossVul~\cite{crossvul}, DiverseVul~\cite{diversevul}, and MegaVul~\cite{megavul} span multiple programming languages or CWE types. However, their labeling approach, which tags pre-patch functions as vulnerable, can result in false positives by mislabeling unrelated code changes.
To address these issues, PrimeVul~\cite{primevul} combines four major VD datasets and applies refined labeling rules, improving labeling accuracy by 26\%–67\%. Given its enhanced diversity and accuracy, we construct our evaluation dataset based on PrimeVul.

\subsubsection{Dataset Partitioning}\label{training_setting}

In real-world applications, VD models should learn from historical vulnerability-related samples to detect future ones. However, most prior studies~\cite{yinxin, diversevul, ni1, ReVeal, linevul} randomly split datasets (8:1:1 ratio), risking data leakage. For instance, a patched (non-vulnerable) function may appear in training, while its pre-patch (vulnerable) version is in testing (Case 1). 
Additionally, similar vulnerability fixes within the same commit (e.g., QEMU 902b27d~\cite{QEMU_commit}
) may be distributed across different data splits, potentially causing the similar patterns learned during training to reappear in the evaluation set (Case 2). To mitigate this, it is necessary to split the dataset based on commit date~\cite{primevul}. 

Furthermore, random splitting could lead to the distribution of different vulnerability types being misaligned with real-world scenarios, resulting in biased conclusions (e.g., Table 7 in DiverseVul~\cite{diversevul}). For instance, some common CWE types may be underrepresented in the training set, which hinders the model from learning relevant vulnerability patterns. Conversely, certain rare CWE types might be mostly allocated to the training set, producing evaluation results on the sparse test data that do not accurately reflect the model’s real-world performance.

To overcome these limitations, we propose a fine-grained data partitioning method that first groups the data by CWE type, then sorts the data by commit date, and finally partitions it within each CWE type according to an 8:1:1 ratio. Our partitioning method not only preserves the original distribution of vulnerability types in the evaluation dataset but also ensures that the model learns relatively sufficient common vulnerability patterns during the training phase. We examine our reconstructed dataset and find no data leakage for Case 1 and 2. This confirms that our approach effectively minimizes data leakage while enabling more accurate evaluation of PLMs’ detection capabilities across individual CWEs.

\subsection{Self-Collected Dataset}\label{new}

Due to potential overlap between the evaluation data for VD and the pre-training data for PLMs (e.g., both collected from same GitHub commits)~\cite{quality}, some studies have independently curated holdout test sets to avoid data leakage and minimize bias in evaluation. However, these datasets often exhibit limitations in terms of data scales, project diversity, and coverage of CWE types. For instance, Ullah et al.~\cite{ullah} curated a dataset consisting of only 228 samples, covering 8 CWE types from 4 projects, which may limit the generalizability of their findings. 

\begin{figure}[h]\centering
\includegraphics[width=0.45\textwidth]{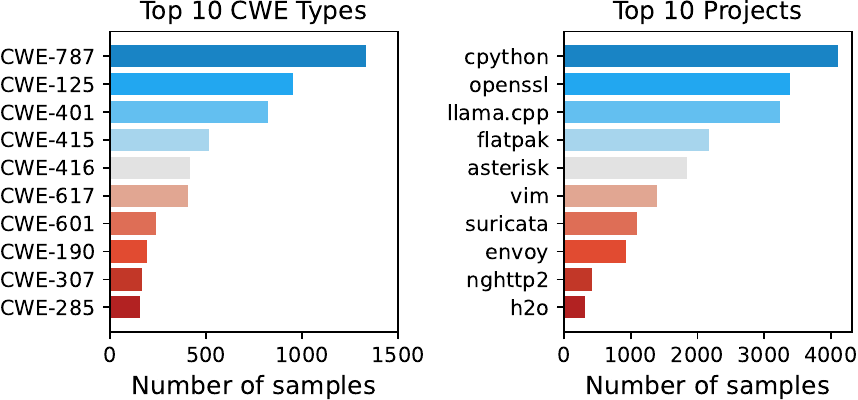}
\caption{Distribution of top 10 CWEs in self-collected data} \label{fig:self-collected}
\vspace{-0.2in}
\Description{Empty}
\end{figure}

To this end, we curate a new VD evaluation dataset by collecting all C/C++ vulnerability-related GitHub commits from NVD with commit dates spanning the period from October 2023 to October 2024, which postdate the pre-training cutoff dates of PLMs evaluated in this study. For each commit, we record detailed metadata about the vulnerability and project information. We segment functions and label them following the prior work~\cite{bigvul, primevul}: changed functions before patching are labeled as vulnerable, while changed functions after patching and unchanged functions are labeled as non-vulnerable. Ultimately, we obtain a evaluation dataset comprising 25,536 functions, including 646 vulnerable functions and 24,890 non-vulnerable functions, spanning 99 projects and 28 CWE types. Fig.~\ref{fig:self-collected} shows the distribution of top 10 vulnerability types of data in our self-collected dataset. Over half of the CWE types are listed among the 2024 top 25 most dangerous software weaknesses~\cite{cwe2024top25}, demonstrating the comprehensiveness of our evaluated data. Note that our automated dataset construction can be easily extended to collect and process the latest patch commits from NVD.

\subsection{Statistics of the Datasets Used in Evaluations}

\begin{table}[h]
  \caption{Statistics of the datasets used in evaluations}
  \label{tab:01statistics}
  \centering
\resizebox{\columnwidth}{!}{
\begin{tabular}{ccccc}
\toprule
\textbf{Dataset}                                                          & \textbf{Used for}                                                                                            & \textbf{\# Vulnerable}                                                   & \textbf{\# Non-Vulnerable}                                                     & \textbf{\# All}                                                                 \\
\midrule
\begin{tabular}[c]{@{}c@{}}Reconstructed\\ PrimeVul\end{tabular} & \begin{tabular}[c]{@{}c@{}}Imbalanced Training\\ Balanced Training\\ Validation\\ Test\end{tabular} & \begin{tabular}[c]{@{}c@{}}5431\\ 5431\\ 678\\ 694\end{tabular} & \begin{tabular}[c]{@{}c@{}}179489\\ 5431\\ 22434\\ 22450\end{tabular} & \begin{tabular}[c]{@{}c@{}}184920\\ 10862\\ 23112\\ 23144\end{tabular} \\
\midrule
Self-collected                                                   & Test                                                                                                & 646                                                             & 24890                                                                 & 25536                                                                  \\
\bottomrule
\end{tabular}
}
\end{table}

To investigate the performance of PLMs under different training and test settings (Section~\ref{sec:RQ2}), we construct a balanced training set based on our reconstructed training set, which by default is imbalanced. Specifically, we adjust the ratio of vulnerable to non-vulnerable functions in the training set to 1:1 through random undersampling, while keeping the original validation and test sets unchanged. The self-collected dataset serves as an out-of-distribution test set to evaluate generalizability while avoiding data leakage (Section~\ref{sec:RQ2} and~\ref{RQ1}). The statistics for the datasets used in our evaluation are provided in Table~\ref{tab:01statistics}.

\subsection{Evaluated Models}

To ensure comprehensive evaluations, we select a set of representative PLMs covering encoder-based (CodeBERT, GraphCodeBERT, UniXCoder, PDBERT), decoder-based (CodeLlama-7B, 13B and 34B, DeepSeek-Coder-6.7B and 33B, StarChat-$\beta$-16B, WizardCoder-15B and 33B, Mistral-7B, GPT-3.5, GPT-4, GPT-4o Mini, and GPT-5 Mini), and encoder-decoder based model architectures (Code T5). Table~\ref{tab:models} and Appendix~\ref{sec:models} show the details of models evaluated in this work.

The rationale for selecting the above PLMs is that all of them are oriented toward code-specific tasks or have been fed on a substantial amount of code data during pre-training. This endows them with a foundational understanding of code, which proves advantageous when applying them to downstream VD tasks. 




\begin{table}[h]
  \caption{Models evaluated in this work}
  \label{tab:models}
  \centering
\resizebox{\columnwidth}{!}{
\begin{tabular}{ccccc}
\toprule
\begin{tabular}[c]{@{}c@{}}\textbf{Model}\\ \textbf{Category}\end{tabular}     & \begin{tabular}[c]{@{}c@{}}\textbf{Model}\\ \textbf{API}\end{tabular}                                                                                                                                                                                                                            & \begin{tabular}[c]{@{}c@{}}\textbf{Parameter}\\ \textbf{Size}\end{tabular}                                   & \begin{tabular}[c]{@{}c@{}}\textbf{Max Position} \\ \textbf{Embeddings}\end{tabular}                        & \begin{tabular}[c]{@{}c@{}}\textbf{Knowledge} \\ \textbf{Cutoff Date}\end{tabular}                                                                                                       \\ \midrule
SLMs                                                         & \begin{tabular}[c]{@{}c@{}}codebert-base\\ graphcodebert-base\\ unixcoder-base-nine\\ codet5-base\\ pdbert-base\end{tabular}                                                                                                                                                   & \begin{tabular}[c]{@{}c@{}}125M\\ 125M\\ 125M\\ 220M\\ 125M\end{tabular}                   & \begin{tabular}[c]{@{}c@{}}512\\ 512\\ 512\\ 512\\ 512\end{tabular}                       & \begin{tabular}[c]{@{}c@{}}\textless Jul 2020\\  \textless Feb 2021\\ \textless Aug 2021\\ \textless Sep 2021\\ \textless Nov 2023\end{tabular}                        \\ \midrule
\begin{tabular}[c]{@{}c@{}}Open-source\\ LLMs\end{tabular}   & \begin{tabular}[c]{@{}c@{}}deepseek-coder-6.7b-instruct\\ deepseek-coder-33b-instruct\\ Mistral-7B-Instruct-v0.1\\ CodeLlama-7b-Instruct-hf\\ CodeLlama-13b-Instruct-hf\\ CodeLlama-34b-Instruct-hf\\ WizardCoder-15B-V1.0\\ WizardCoder-33B-V1.1\\ starchat-beta\end{tabular} & \begin{tabular}[c]{@{}c@{}}6.7B\\ 33B\\ 7B\\ 7B\\ 13B\\ 34B\\ 15B\\ 33B\\ 16B\end{tabular} & \begin{tabular}[c]{@{}c@{}}16K\\ 16K\\ 32K\\ 16K\\ 16K\\ 16K\\ 8K\\ 16K\\ 8K\end{tabular} & \begin{tabular}[c]{@{}c@{}}Mar 2023\\ Mar 2023\\ Aug 2021\\ Jan 2023\\ Jan 2023\\ Jan 2023\\ \textless Jun 2023\\ \textless Jun 2023\\ \textless Jun 2023\end{tabular} \\ \midrule
\begin{tabular}[c]{@{}c@{}}Closed-source\\ LLMs\end{tabular} & \begin{tabular}[c]{@{}c@{}}gpt-3.5-turbo-0125\\ gpt-4-32k\\ gpt-4o-mini\\ gpt-5-mini\end{tabular}                                                                                                                                                                                           & \begin{tabular}[c]{@{}c@{}}7B\\ -\\ -\end{tabular}                                         & \begin{tabular}[c]{@{}c@{}}16K\\ 32K\\ 128K\\ 400K\end{tabular}                                  & \begin{tabular}[c]{@{}c@{}}Sep 2021\\ Sep 2021\\ Oct 2023\\ Jun 2024\end{tabular}                                                                                                 \\ \bottomrule
\end{tabular}
}
\end{table}

\subsection{Prompt Design}\label{sec:PDesign}

\noindent\textbf{Prompt Settings.} We evaluate the performance of LLMs in VD under two prompt settings: (1) zero-shot prompting: directly asking the LLM if the function is vulnerable without providing any additional information; (2) few-shot in-context learning (ICL): providing the LLM with 4 shots (2 randomly selected pre-patched and post-patched function pairs each time) beforehand, with each shot including a query and its ground truth answer, followed by asking the LLM whether a target function is vulnerable. 

Under each prompt setting, we categorize the prompts into four types: \ding{172} raw code, \ding{173} flattened AST, \ding{174} code with API calls, and \ding{175} code with data flow. In 
\ding{173}-\ding{175}, the prompt contains not only raw code but also structural information within the code, to explore if this assists the LLM in understanding the code and its vulnerability patterns. We use Tree-sitter~\cite{treesitter} to extract the structural information from the code and serialize it into plain text.

For \ding{173}, after obtaining the AST of code, we adopt UniXCoder~\cite{UniXcoder}'s AST mapping algorithm to generate flattened AST.
For example, the code line {\small{\texttt{c=a+b;}}} is translated to:

\begin{tcolorbox}[breakable,width=\linewidth,colback=cyan!10, colframe=white,boxrule=0pt, left=1mm, right=1mm, top=1mm, bottom=1mm]
\noindent\textbf{Flattened AST}:\\
\textless AST\#expression\_statement\#Left\textgreater\newline
\text{\hspace{5pt}}\textless AST\#assignment\_expression\#Left\textgreater\ c = \newline
\text{\hspace{10pt}}\textless AST\#binary\_expression\#Left\textgreater\ a + b\newline
\text{\hspace{10pt}}\textless AST\#binary\_expression\#Right\textgreater\newline
\text{\hspace{5pt}}\textless AST\#assignment\_expression\#Right\textgreater\ ; \newline
\textless AST\#expression\_statement\#Right\textgreater
\end{tcolorbox}

For \ding{174}, we traverse the AST and collect nodes of type ``call\-expression''. Following prior work~\cite{graph_prompt}, we describe the flow of API call using the following template:

\begin{tcolorbox}[breakable,width=\linewidth,colback=cyan!10, colframe=white,boxrule=0pt, left=1mm, right=1mm, top=1mm, bottom=1mm]
\noindent\textbf{API call}:
The program first calls \textless node\_0\textgreater, ... , then calls \textless node\_1\textgreater, ... , and finally calls \textless node\_n\textgreater.
\end{tcolorbox}

For \ding{175}, inspired by GraphCodeBERT~\cite{graphcodebert}, which incorporates data flow during pre-training to enhance code understanding, we explore the effect of introducing data flow descriptions during inference. We extract data flow between variable nodes in the format (VAR, $p_i$, comesFrom, [VAR], [$p_j$]), indicating that the variable VAR at position $p_i$ originates from another at $p_j$. While~\cite{graph_prompt} describes this as ``the data value of VAR at the $p_i$th token comes from data at the $p_j$th token", using absolute token positions could misalign with LLM tokenization, leading to incorrect interpretation. To 
ensure that the LLM accurately identifies tokens and understands the data flow, we use the relative position of the node VAR within the function and translate the above example as below:

\begin{tcolorbox}[breakable,width=\linewidth,colback=cyan!10, colframe=white,boxrule=0pt, left=1mm, right=1mm, top=1mm, bottom=1mm]
\noindent\textbf{Data flow}:
The 2nd VAR comes from the 1st VAR ...
\end{tcolorbox}

\noindent\textbf{Prompt Templates.} To guide LLMs in making better predictions, our prompts are composed of two components: system role and user content. For the system role, the prompts begin with the instruction: ``You are a code security expert who excels at detecting vulnerabilities". The user content starts with the question: ``Is the following function vulnerable? Please answer Yes or No", followed by the input function. Additionally, we adopt the default chat templates of LLMs by calling \texttt{\small{tokenizer.apply\_chat\_template()}}~\cite{huggingface_chat_templates} to align with their instruction data format.

\begin{table*}[h]
  \caption{Comparison of PLMs under imbalanced and balanced training settings}
  \label{tab:01balanced}
  \centering
\resizebox{0.8\textwidth}{!}{

\begin{tabular}{cccccccc}

\toprule
\multirow{2}{*}{\textbf{Settings}}                                     & \multirow{2}{*}{\textbf{Models}}                                                                        & \multicolumn{6}{c}{\textbf{Imbalanced Testing on Reconstructed Test Set\hspace{0.2em}\textbar\hspace{0.2em}Self-collected Dataset}}                                                                                                                                                                                                                                                                                                                                                                                                                                                                                                                                                                                                                                                                                                                                                                                                                                                                                                                                                                                                                                                                                                                                                                                                                                                                                                                                                                \\ \cline{3-8} 
                                                              &                                                                                                & \multicolumn{1}{c}{\textbf{Accuracy}}                                                                                                                                                                                                                  & \multicolumn{1}{c}{\textbf{Balanced Accuracy}}                                                                                                                                                                                                         & \multicolumn{1}{c}{\textbf{F1}}                                                                                                                                                                                                                        & \multicolumn{1}{c}{\textbf{Precision}}                                                                                                                                                                                                                 & \multicolumn{1}{c}{\textbf{Recall}}                                                                                                                                                                                                                    & \textbf{TNR}                                                                                                                                                                                                                       \\ \midrule
\begin{tabular}[c]{@{}c@{}}Imbalanced\\ Training\end{tabular} & \begin{tabular}[c]{@{}c@{}}CodeBERT\\ UniXCoder\\ GraphCodeBERT\\ PDBERT\\ CodeT5\end{tabular} & \multicolumn{1}{c}{\begin{tabular}[c]{@{}c@{}}96.24\hspace{0.2em}\textbar\hspace{0.2em}97.18\\ 96.40\hspace{0.2em}\textbar\hspace{0.2em}97.31\\ 96.34\hspace{0.2em}\textbar\hspace{0.2em}97.11\\ 96.80\hspace{0.2em}\textbar\hspace{0.2em}97.57\\ 96.51\hspace{0.2em}\textbar\hspace{0.2em}97.35\end{tabular}} & \multicolumn{1}{c}{\begin{tabular}[c]{@{}c@{}}54.91\hspace{0.2em}\textbar\hspace{0.2em}51.73\\ 53.46\hspace{0.2em}\textbar\hspace{0.2em}50.82\\ 55.03\hspace{0.2em}\textbar\hspace{0.2em}51.25\\ 57.92\hspace{0.2em}\textbar\hspace{0.2em}51.93\\ 53.03\hspace{0.2em}\textbar\hspace{0.2em}50.92\end{tabular}} & \multicolumn{1}{c}{\begin{tabular}[c]{@{}c@{}}14.87\hspace{0.2em}\textbar\hspace{0.2em}6.49\\ 11.48\hspace{0.2em}\textbar\hspace{0.2em}3.37\\ 15.37\hspace{0.2em}\textbar\hspace{0.2em}4.90\\ 23.69\hspace{0.2em}\textbar\hspace{0.2em}7.45\\ 10.43\hspace{0.2em}\textbar\hspace{0.2em}3.70\end{tabular}}      & \multicolumn{1}{c}{\begin{tabular}[c]{@{}c@{}}23.17\hspace{0.2em}\textbar\hspace{0.2em}20.00\\ 21.86\hspace{0.2em}\textbar\hspace{0.2em}18.18\\ 25.00\hspace{0.2em}\textbar\hspace{0.2em}14.73\\ 41.52\hspace{0.2em}\textbar\hspace{0.2em}100.0\\ 22.71\hspace{0.2em}\textbar\hspace{0.2em}22.81\end{tabular}} & \multicolumn{1}{c}{\begin{tabular}[c]{@{}c@{}}10.95\hspace{0.2em}\textbar\hspace{0.2em}3.87\\ 7.78\hspace{0.2em}\textbar\hspace{0.2em}1.86\\ 11.10\hspace{0.2em}\textbar\hspace{0.2em}2.94\\ 16.57\hspace{0.2em}\textbar\hspace{0.2em}3.87\\ 6.77\hspace{0.2em}\textbar\hspace{0.2em}2.01\end{tabular}}        & \begin{tabular}[c]{@{}c@{}}\cellcolor{red!30}98.88\hspace{0.2em}\textbar\hspace{0.2em}\cellcolor{red!30}99.60\\ \cellcolor{red!30}99.14\hspace{0.2em}\textbar\hspace{0.2em}\cellcolor{red!30}99.78\\ \cellcolor{red!30}98.97\hspace{0.2em}\textbar\hspace{0.2em}\cellcolor{red!30}99.56\\ \cellcolor{red!30}99.28\hspace{0.2em}\textbar\hspace{0.2em}\cellcolor{red!30}100.0\\ \cellcolor{red!30}99.29\hspace{0.2em}\textbar\hspace{0.2em}\cellcolor{red!30}99.82\end{tabular} \\ \midrule
\begin{tabular}[c]{@{}c@{}}Balanced\\ Training\end{tabular}   & \begin{tabular}[c]{@{}c@{}}CodeBERT\\ UniXCoder\\ GraphCodeBERT\\ PDBERT\\ CodeT5\end{tabular} & \multicolumn{1}{c}{\begin{tabular}[c]{@{}c@{}}75.34\hspace{0.2em}\textbar\hspace{0.2em}82.44\\ 71.75\hspace{0.2em}\textbar\hspace{0.2em}77.91\\ 69.82\hspace{0.2em}\textbar\hspace{0.2em}73.48\\ 76.91\hspace{0.2em}\textbar\hspace{0.2em}81.47\\ 73.94\hspace{0.2em}\textbar\hspace{0.2em}78.34\end{tabular}} & \multicolumn{1}{c}{\begin{tabular}[c]{@{}c@{}}\cellcolor{green!20}73.11\hspace{0.2em}\textbar\hspace{0.2em}\cellcolor{green!10}67.02\\ \cellcolor{green!20}74.62\hspace{0.2em}\textbar\hspace{0.2em}\cellcolor{green!20}68.61\\ \cellcolor{green!20}73.48\hspace{0.2em}\textbar\hspace{0.2em}\cellcolor{green!10}67.62\\ \cellcolor{green!30}78.33\hspace{0.2em}\textbar\hspace{0.2em}\cellcolor{green!30}71.27\\ \cellcolor{green!20}72.32\hspace{0.2em}\textbar\hspace{0.2em}\cellcolor{green!10}67.86\end{tabular}} & \multicolumn{1}{c}{\begin{tabular}[c]{@{}c@{}}14.68\hspace{0.2em}\textbar\hspace{0.2em}12.77\\ 14.15\hspace{0.2em}\textbar\hspace{0.2em}11.87\\ 13.33\hspace{0.2em}\textbar\hspace{0.2em}10.49\\ 17.18\hspace{0.2em}\textbar\hspace{0.2em}14.18\\ 13.98\hspace{0.2em}\textbar\hspace{0.2em}11.72\end{tabular}} & \multicolumn{1}{c}{\begin{tabular}[c]{@{}c@{}}8.19\hspace{0.2em}\textbar\hspace{0.2em}7.30\\ 7.79\hspace{0.2em}\textbar\hspace{0.2em}6.60\\ 7.29\hspace{0.2em}\textbar\hspace{0.2em}5.74\\ 9.62\hspace{0.2em}\textbar\hspace{0.2em}8.03\\ 7.76\hspace{0.2em}\textbar\hspace{0.2em}6.53\end{tabular}}           & \multicolumn{1}{c}{\begin{tabular}[c]{@{}c@{}}70.75\hspace{0.2em}\textbar\hspace{0.2em}50.77\\ 77.67\hspace{0.2em}\textbar\hspace{0.2em}58.82\\ 77.38\hspace{0.2em}\textbar\hspace{0.2em}61.46\\ 79.83\hspace{0.2em}\textbar\hspace{0.2em}60.53\\ 70.61\hspace{0.2em}\textbar\hspace{0.2em}56.81\end{tabular}} & \begin{tabular}[c]{@{}c@{}}75.48\hspace{0.2em}\textbar\hspace{0.2em}83.27\\ 71.57\hspace{0.2em}\textbar\hspace{0.2em}78.40\\ 69.59\hspace{0.2em}\textbar\hspace{0.2em}73.79\\ 76.82\hspace{0.2em}\textbar\hspace{0.2em}82.01\\ 74.04\hspace{0.2em}\textbar\hspace{0.2em}78.90\end{tabular} \\ \bottomrule
\end{tabular}
}

\end{table*}

\subsection{Metrics}

We consider multiple metrics to comprehensively demonstrate the models’ capabilities, formulated as follows:
\textbf{Accuracy} = $(TP + TN)/(TP + TN + FP + FN)$,
\textbf{Recall} = $TP/(TP + FN)$,
\textbf{Precision} = $TP$/$(TP + FP)$,
\textbf{True Negative Rate (TNR)} = $TN/(TN + FP)$,
\textbf{F1} = $2 \cdot (\text{Precision}\cdot\text{Recall})/(\text{Precision} + \text{Recall})$,
\textbf{Balanced Accuracy} = $(\text{Recall} + \text{TNR}))/2$. Since we expect VD models to exhibit both low false positive and false negative rates in practice, following previous work~\cite{benjamin}, we use balanced accuracy as our main metric to balance true positive predictions and true negative predictions.

\subsection{Implementation}
\subsubsection{Evaluation Framework}
Given the variations in implementation across existing empirical studies on PLMs for VD~\cite{yinxin, primevul, benjamin2023, benjamin, ullah, pdbert}, we develop a comprehensive evaluation framework for PLMs in VD, building upon ~\cite{primevul, benjamin}. This framework is designed to be easily extensible to new datasets, PLMs, model adaptation techniques, with the goal of enhancing the accuracy and fairness of PLM evaluations for VD. 
Specifically, for open-source PLMs, we utilize the Hugging Face transformers library~\cite{hf} to load configurations, tokenizers, and base models. For SLMs, we perform full fine-tuning. For LLMs, we use LoRA to fine-tune the models and incorporate DeepSpeed ZeRO-3 along with the Accelerate module to improve training efficiency. For GPT models, we use the Azure OpenAI API for inference. 

All experiments are conducted on a computational node equipped with multiple NVIDIA A100 GPUs (80 GB VRAM each), 64 CPU cores, and 512 GB of RAM. For GPTs, approximately 38 million tokens are processed per model, with a total cost of \$2,400.

\subsubsection{Hyperparameters}\label{sec:hparam}

\texttt{\small{RobertaClassificationHead}}~\cite{hf} is used as classifier when fine-tuning SLMs, which comprises a feedforward neural network with two linear layers (hidden\_size=768), dropout regularization (rate=0.1), and a tanh activation function. The AdamW optimizer~\cite{adamw} is employed along with a linear learning rate schedule with a warm-up ratio set to 10\% of total training steps. When fine-tuning LLMs, we use \texttt{\small{AutoModelForSequence\-Classification}} to load base model and add LORA adapters to all linear layers of base model. We configure LoRA adapter with a rank of 64, an alpha of 16, and 0.05 dropout rate. For all experiments, we use a batch size of 32, 10 epochs, and a learning rate of 2e-5.

When prompting LLMs, we set the top-p value to 0.9, the temperature to 0. For GPT-5 Mini, we keep the temperature to default 1.0 as it cannot be modified. The reasoning effort is set to medium, and the verbosity is low because we only focus on the correctness of the binary label prediction. These settings are chosen to ensure that the model produces concise and deterministic answers.

\section{Research Questions (RQs) and Findings}

\subsection{RQ1: How do PLMs perform across various training and testing configurations?
}\label{sec:RQ2}


Due to inconsistent experimental settings in existing VD research, it is difficult to compare evaluation results across studies, often leading to conflicting conclusions. For example, some studies fine-tune PLMs on imbalanced training sets and evaluate them on imbalanced test sets~\cite{primevul, diversevul}, while others fine-tune on balanced training sets and evaluate on either imbalanced~\cite{yinxin} or balanced test sets~\cite{davidlo}. To investigate the impact of different experimental settings on model training and testing, we fine-tune Code SLMs under both imbalanced and balanced training conditions (LLMs are excluded because they share the same conclusion as those of Code SLMs). We then evaluate their generalizability using two imbalanced test sets with different distributions. We deliberately avoid using balanced test settings, as real-world security data is inherently imbalanced. Evaluating a model, regardless of its training setup, solely on balanced data risks significantly overestimating its precision.


From Table~\ref{tab:01balanced}, we observe that under imbalanced training settings, the average TNR of Code SLMs on both test sets exceeds 99\%, whereas the average recall remains below 10\%. This discrepancy is attributed to the overwhelming number of non-vulnerable functions compared to vulnerable ones, which biases the fine-tuned Code SLMs toward predicting negatives. Under balanced training settings, the average TNR of Code SLMs on the reconstructed test set is 73.5\%, the average recall is 75.3\%, and the average balanced accuracy is 74.4\%. These suggest that Code SLMs fine-tuned under balanced settings are able to fairly learn the representations of both vulnerable and non-vulnerable functions, thereby achieving a better trade-off between recall and TNR. In particular, 
we observe that the F1 score tends to favor models trained on highly imbalanced data, creating a misleading impression that these models perform better under imbalanced training settings. In addition, all PLMs show a low precision, which is because the imbalanced distribution of real-world vulnerability data (non-vulnerable data is far more than vulnerable data) results in the number of FPs being much larger than the number of TPs. This also explains why we chose balanced accuracy as our main metric; it considers both recall and TNR, so any bias toward either class will equally affect the result.

Among Code SLMs fine-tuned under balanced training settings, PD\-BERT performs the best with its balanced accuracy averaging about 3.2\% to 4.7\% higher than that of the others on both test sets. UniXCoder and GraphCodeBERT follow behind, while CodeBERT exhibits the lowest performance. PDBERT's superior performance is attributed to its incorporation of both control and data dependency prediction tasks during pre-training, which enables the model to capture vulnerability-related patterns by analyzing dependencies within the code, thereby enhancing its detection ability. In contrast, UniXCoder and GraphCodeBERT, which are exposed to only single aspects of data flow or AST during pre-training, show somewhat lower performance. Without learning code structure during pre-training, CodeBERT presents limited ability to accurately distinguish between vulnerable and non-vulnerable functions.

In terms of generalizability, we observe that under balanced training settings, although SLMs perform well on the reconstructed test set, their performance drops significantly on the self-collected dataset, with average balanced accuracy decreasing by about 5.9\%. These results suggest that the generalization ability of code language models in real-world scenarios still requires improvement.
\begin{tcolorbox}[breakable,width=\linewidth,colback=cyan!10, colframe=white,boxrule=0pt, left=1mm, right=1mm, top=1mm, bottom=1mm]
\noindent\textbf{Answer-1}: (1) Selecting appropriate experimental settings is essential for accurately evaluating the capabilities of PLMs for VD. Fine-tuning PLMs under balanced training settings helps mitigate prediction bias. Evaluating PLMs on imbalanced test sets ensures a realistic performance assessment in practical scenarios. (2) Among existing Code SLMs, PDBERT achieves the best performance, demonstrating that guiding models to learn code dependency relationships during pre-training effectively enhances VD. (3) Evaluating fine-tuned PLMs on test data sourced from the same distribution as the training data leads to overestimated results that fail to reflect the models’ real ability for VD in real-world scenarios. 
\end{tcolorbox}
\begin{table*}[h]
\caption{Performance comparison of Code SLMs and LLMs on the self-collected evaluation dataset}
\label{tab:GeneralComparison}
\footnotesize
  \centering
\begin{tabular}{cccccccc}
\toprule
                                                                                   & \multicolumn{1}{c}{}                                                                                                                                                                                                        & \multicolumn{1}{c}{\textbf{Accuracy}}                                                                                                      & \multicolumn{1}{c}{\begin{tabular}[c]{@{}c@{}}\textbf{Balanced}\\ \textbf{Accuracy}\end{tabular}}                                                   & \multicolumn{1}{c}{\textbf{F1}}                                                                                                            & \multicolumn{1}{c}{\textbf{Precision}}                                                                                            & \multicolumn{1}{c}{\textbf{Recall}}                                                                                                        & \textbf{TNR}                                                                                                           \\ \midrule
\multicolumn{8}{l}{\textbf{Fine Tuning-based Evaluation}}                                                                                                                                                                                                                                                                                                                                                                                                                                                                                                                                                                                                                                                                                                                                                                                                                                                                                                                                                                                                                                                                \\ \midrule
\multicolumn{1}{c}{Code SLMs}                                                    & \multicolumn{1}{c}{\begin{tabular}[c]{@{}c@{}}CodeBERT\\ UniXCoder\\ GraphCodeBERT\\ PDBERT\\ CodeT5\end{tabular}}                                                                                                          & \multicolumn{1}{c}{\begin{tabular}[c]{@{}c@{}}82.44\\ 77.91\\ 73.48\\ 81.47\\ 78.34\end{tabular}}                                 & \multicolumn{1}{c}{\begin{tabular}[c]{@{}c@{}}\cellcolor{green!5}67.02\\ \cellcolor{green!10}68.61\\ \cellcolor{green!5}67.62\\ \cellcolor{green!30}71.27\\ \cellcolor{green!5}67.86\end{tabular}}                                 & \multicolumn{1}{c}{\begin{tabular}[c]{@{}c@{}}12.77\\ 11.87\\ 10.49\\ 14.18\\ 11.72\end{tabular}}                                 & \multicolumn{1}{c}{\begin{tabular}[c]{@{}c@{}}7.30\\ 6.60\\ 5.74\\ 8.03\\ 6.53\end{tabular}}                             & \multicolumn{1}{c}{\begin{tabular}[c]{@{}c@{}}50.77\\ 58.82\\ 61.46\\ 60.53\\ 56.81\end{tabular}}                                 & \begin{tabular}[c]{@{}c@{}}83.27\\ 78.40\\ 73.79\\ 82.01\\ 78.90\end{tabular}                                 \\ \midrule
\multicolumn{1}{c}{LLMs}                                                         & \multicolumn{1}{c}{\begin{tabular}[c]{@{}c@{}}DeepSeek-Coder-6.7B\\ DeepSeek-Coder-34B\\ Mistral-7B\\ CodeLlama-7B\\ CodeLlama-13B\\ CodeLlama-34B\\ WizardCoder-15B\\ WizardCoder-34B\\ Starchat-$\beta$-16B\end{tabular}} & \multicolumn{1}{c}{\begin{tabular}[c]{@{}c@{}}73.94\\ 76.11\\ 75.94\\ 75.83\\ 78.30\\ 79.08\\ 78.06\\ 77.58\\ 80.63\end{tabular}} & \multicolumn{1}{c}{\begin{tabular}[c]{@{}c@{}}\cellcolor{green!10}68.99\\ \cellcolor{green!20}70.48\\ \cellcolor{green!10}68.06\\ \cellcolor{green!20}69.51\\ \cellcolor{green!10}68.36\\ \cellcolor{green!5}65.67\\ \cellcolor{green!5}65.45\\ \cellcolor{green!10}68.90\\ \cellcolor{green!5}66.47\end{tabular}} & \multicolumn{1}{c}{\begin{tabular}[c]{@{}c@{}}11.02\\ 12.03\\ 11.16\\ 11.63\\ 11.89\\ 11.09\\ 10.74\\ 11.88\\ 11.87\end{tabular}} & \multicolumn{1}{c}{\begin{tabular}[c]{@{}c@{}}6.03\\ 6.63\\ 6.16\\ 6.41\\ 6.63\\ 6.21\\ 5.99\\ 6.60\\ 6.70\end{tabular}} & \multicolumn{1}{c}{\begin{tabular}[c]{@{}c@{}}63.78\\ 64.55\\ 59.75\\ 62.85\\ 57.89\\ 51.55\\ 52.17\\ 59.75\\ 51.55\end{tabular}} & \begin{tabular}[c]{@{}c@{}}74.21\\ 76.41\\ 76.36\\ 76.17\\ 78.83\\ 79.80\\ 78.73\\ 78.04\\ 81.38\end{tabular} \\ \midrule
\multicolumn{8}{l}{\textbf{Prompting-based Evaluation}}                                                                                                                                                                                                                                                                                                                                                                                                                                                                                                                                                                                                                                                                                                                                                                                                                                                                                                                                                                                                                                                                  \\ \midrule
\multicolumn{1}{c}{\begin{tabular}[c]{@{}c@{}}Open-sourced\\ LLMs\end{tabular}}  & \multicolumn{1}{c}{\begin{tabular}[c]{@{}c@{}}DeepSeek-Coder-6.7B\\ DeepSeek-Coder-34B\\ Mistral-7B\\ CodeLlama-7B\\ CodeLlama-13B\\ CodeLlama-34B\\ WizardCoder-15B\\ WizardCoder-34B\\ Starchat-$\beta$-16B\end{tabular}} & \multicolumn{1}{c}{\begin{tabular}[c]{@{}c@{}}50.72\\ 63.92\\ 32.68\\ 2.61\\ 2.53\\ 14.34\\ 2.50\\ 52.78\\ 9.19\end{tabular}}     & \multicolumn{1}{c}{\begin{tabular}[c]{@{}c@{}}\cellcolor{red!10}53.08\\ \cellcolor{green!5}61.44\\ \cellcolor{red!10}54.16\\ \cellcolor{red!30}49.89\\ \cellcolor{red!30}50.01\\ \cellcolor{red!20}51.61\\ \cellcolor{red!30}49.00\\ \cellcolor{red!5}58.36\\ \cellcolor{red!30}50.25\end{tabular}} & \multicolumn{1}{c}{\begin{tabular}[c]{@{}c@{}}5.40\\ 7.62\\ 5.46\\ 4.92\\ 4.93\\ 5.09\\ 4.84\\ 6.44\\ 4.95\end{tabular}}          & \multicolumn{1}{c}{\begin{tabular}[c]{@{}c@{}}2.84\\ 4.07\\ 2.83\\ 2.52\\ 2.53\\ 2.62\\ 2.48\\ 3.39\\ 2.54\end{tabular}} & \multicolumn{1}{c}{\begin{tabular}[c]{@{}c@{}}55.57\\ 58.82\\ 76.78\\ 99.69\\ 100.0\\ 90.87\\ 97.99\\ 64.24\\ 93.50\end{tabular}}   & \begin{tabular}[c]{@{}c@{}}50.59\\ 64.05\\ 31.54\\ 0.10\\ 0.00\\ 12.35\\ 0.02\\ 52.49\\ 7.00\end{tabular}     \\ \midrule
\multicolumn{1}{c}{\begin{tabular}[c]{@{}c@{}}Close-sourced\\ LLMs\end{tabular}} & \multicolumn{1}{c}{\begin{tabular}[c]{@{}c@{}}GPT-3.5\\ GPT-4\\ GPT-4o Mini\end{tabular}}                                                                                                                                   & \multicolumn{1}{c}{\begin{tabular}[c]{@{}c@{}}52.17\\ 61.11\\ 51.92\end{tabular}}                                                 & \multicolumn{1}{c}{\begin{tabular}[c]{@{}c@{}}\cellcolor{red!10}52.47\\ \cellcolor{green!5}60.00\\ \cellcolor{green!5}62.67\end{tabular}}                                                 & \multicolumn{1}{c}{\begin{tabular}[c]{@{}c@{}}5.29\\ 7.11\\ 7.22\end{tabular}}                                                    & \multicolumn{1}{c}{\begin{tabular}[c]{@{}c@{}}2.78\\ 3.78\\ 3.80\end{tabular}}                                           & \multicolumn{1}{c}{\begin{tabular}[c]{@{}c@{}}52.79\\ 58.82\\ 73.99\end{tabular}}                                                 & \begin{tabular}[c]{@{}c@{}}52.16\\ 65.17\\ 51.35\end{tabular}                                                 \\ \bottomrule
\end{tabular}
\end{table*}

\subsection{RQ2: How do PLMs of varying parameter scales perform under both fine-tuning and prompt engineering?}\label{RQ1}

Given the strong performance of LLMs in NLP and software engineering, and the efficiency of prompt engineering compared to fine-tuning, most prior studies prefer prompting-based evaluations for VD, leaving comprehensive evaluations of fine-tuned LLMs with parameter sizes up to 34B underexplored. Additionally, Code SLMs, pre-trained on structure-aware code tasks, are expected to be better suited for VD due to their ability to capture complex code structures and dependencies. With significantly fewer parameters than LLMs, Code SLMs can be fully fine-tuned on VD datasets with less computational cost. To this end, we compare Code SLMs and LLMs on VD tasks under both fine-tuning and prompt settings, analyzing the trade-off between performance and efficiency.

Specifically, in prompting-based evaluations, for each LLM, we apply four types of prompts (Section~\ref{sec:PDesign}) under both zero-shot and few-shot settings, and report the best testing results. 
The fine-tuning-based evaluations are conducted in balanced reconstruction training set (best setting shown in Section~\ref{sec:RQ2}). For Code SLMs, we perform full fine-tuning. For LLMs, we apply parameter-efficient fine-tuning (PEFT) using LoRA~\cite{hu2022lora}, which has shown performance comparable to full fine-tuning.
To avoid data leakage and ensure accurate evaluation, all PLMs are tested on our new self-collected dataset containing commits from October 2023, after the pretraining cutoff of all evaluated models.

As shown in Table~\ref{tab:GeneralComparison}, 
the overall performance of fine-tuned PLMs significantly outperform that of prompting-based PLMs by about 14\%. This suggests that, although few-shot examples or code structure and dependency information are provided during prompting to guide LLMs in understanding code, the complexity and diversity of vulnerability patterns still pose substantial challenges. In contrast, fine-tuning PLMs involves feeding the models a large amount of vulnerability-related code and guiding them through supervised classification to distinguish between vulnerable and non-vulnerable code, thereby enhancing their capabilities for VD.

In the prompting-based evaluation, GPT-4o Mini and DeepSeek-Coder-34B perform the best, confirming their strong performance in code-related tasks. As the parameter scale increases, the performance of LLMs also improves, validating that more code data fed during the pre-training phase enhances the models’ general understanding of code. This in turn aids in downstream VD tasks by allowing models to capture code logic and dependency relationships for detecting potential vulnerabilities. Notably, for CodeLlama and some relatively smaller LLMs (e.g., WizardCoder-15B and Starchat-$\beta$-16B), their TNR is below 10\%. As parameter scale increases, this issue is partially mitigated in WizardCoder, with its 34B model showing a 52\% improvement in TNR compared to its 15B counterpart. However, for CodeLlama, despite maintaining a high recall, its TNR remains low. This may be due to the model’s oversensitivity to certain keywords in prompts, leading to unreliability in VD~\cite{benjamin2023}.

In the fine-tuning-based evaluation, PDBERT and DeepSeek-Coder-34B perform best, achieving high recall and TNR, demonstrating strong capability in distinguishing between vulnerable and non-vulnerable code. Codellama-7B and DeepSeek-Coder-6.7 closely follow in performance. Overall, as model size increases, the performance of fine-tuned LLMs (e.g., DeepSeek-Coder, WizardCoder) also improves. However, for CodeLlama, the scaling law appears to break down. There is a dramatic drop in recall accompanied by an increase in TNR. Considering its sensitivity to keywords shown in prompting-based evaluation, the possible reason is that during fine-tuning, the model more aggressively overwrites its general-purpose pre-training knowledge when exposed to a relatively small yet pattern-complex VD dataset.

\begin{figure}[h]\centering
\includegraphics[width=0.45\textwidth]{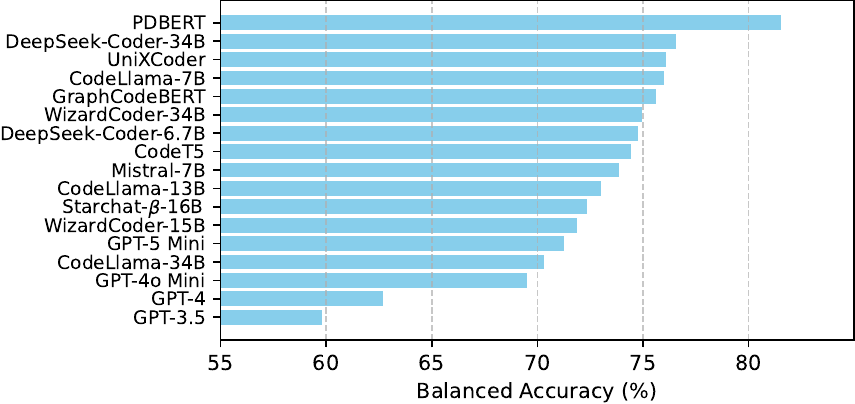}
\caption{Performance comparison of PLMs on the self-collected dataset beyond the GPT-5 Mini cutoff date} \label{fig:new_exp_asiaccs}
\Description{Empty}
\end{figure}

\noindent\textbf{Performance Evaluation of GPT-5 Mini}. Recent advancements in PLMs with reasoning capabilities have shown strong performance in the field of software engineering. To complement this and highlight the timeliness of our findings, we include the SOTA reasoning model GPT-5 Mini, which has demonstrated leading performance in code benchmarks such as SWE-Bench~\cite{swebench}.

As the knowledge cutoff date for GPT-5 Mini is June 2024, we select a subset of the self-collected evaluation data with commit dates starting from July 2024 to ensure accuracy. For a fair comparison, we also report the best balanced accuracy of other PLMs on the same filtered subset. As shown in Fig.~\ref{fig:new_exp_asiaccs}, while GPT-5 Mini shows a relative improvement in performance compared to GPT-4o Mini, it still fails to outperform most fine-tuned PLMs. This once again underscores the importance of fine-tuning for improving PLMs' capabilities for VD.

\begin{tcolorbox}[breakable,width=\linewidth,colback=cyan!10, colframe=white,boxrule=0pt, left=1mm, right=1mm, top=1mm, bottom=1mm]
\noindent\textbf{Answer-2}: (1) Existing research tends to favor applying prompting-based PLMs for VD, while overlooking a comprehensive evaluation of fine-tuned PLMs across parameter scales. While prompt engineering is efficient, its performance on VD tasks falls far short of that achieved through fine-tuning. This highlights the importance of balancing performance and efficiency in practical applications. (2) Scaling law generally holds in both fine-tuning and prompting-based evaluations, but its impact varies across PLMs. (3) Compared to pre-training solely on large code corpora with standard language modeling, incorporating code-specific optimization during pre-training proves more beneficial for achieving strong performance on VD tasks, which demand a deep understanding of code logic and dependencies.
\end{tcolorbox}


\subsection{RQ3: How do PLMs perform across various types of vulnerabilities?}\label{RQ3}

\begin{figure}[h]\centering
\includegraphics[width=0.45\textwidth]{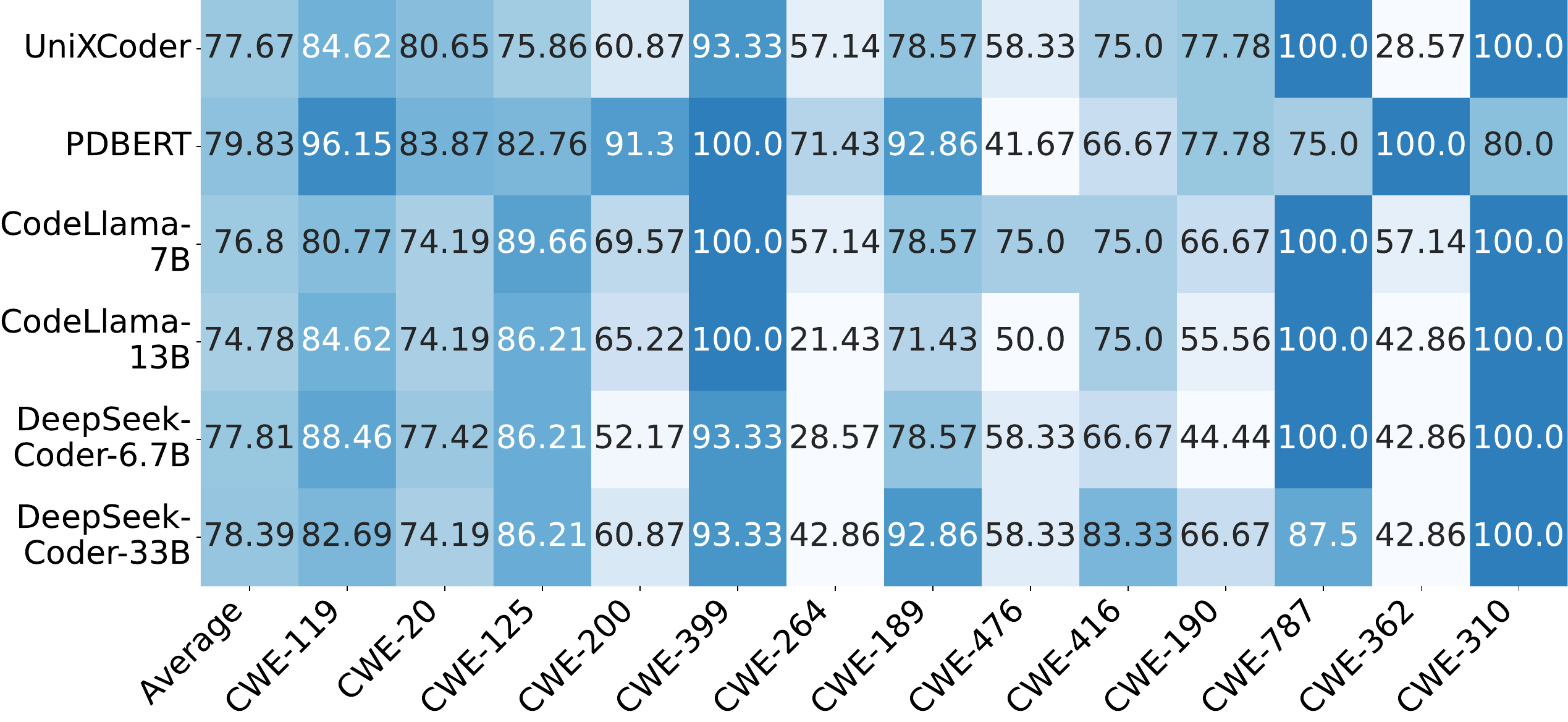}
\caption{PLMs’ recall by vulnerability type\\(ordered by decreasing frequency from left to right)} \label{fig:cwe}
\vspace{-0.1in}
\Description{Empty}
\end{figure}

In this section, we evaluate the ability of PLMs to detect individual CWEs on our reconstructed dataset. To ensure the representativeness of conclusions, in Sections~\ref{RQ3}-\ref{RQ6}, we select representative models that perform well in Table~\ref{tab:GeneralComparison} from different parameter scales. As shown in Fig.~\ref{fig:cwe}, we observe that all PLMs achieve over 70\% recall on average and perform particularly well on specific CWEs, such as CWE-119 (Improper Restriction of Operations within the Bounds of a Memory Buffer), CWE-399 (Resource Management Errors), CWE-189 (Numeric Errors), and CWE-787 (Out-of-bounds Write). Detecting these types of vulnerabilities often involves recognizing the absence of sanity checks within the code context. For example, CWE-119 typically arises when a program fails to validate boundary sizes. PLMs can relatively easily identify that by checking whether memory buffer are properly accessed (e.g., adding boundary checks before sensitive operations).

In particular, PDBERT performs significantly better on CWE-119, CWE-200 (Exposure of Sensitive Information to an Unauthorized Actor), CWE-189, and CWE-362 (Concurrent Execution using Shared Resource with Improper Synchronization). This is attributed to PDBERT’s ability to learn how to capture program dependencies during the pre-training phase, which enables it to detect related vulnerabilities by tracking control and data flow in code.

However, it can be challenging for most PLMs to detect certain types of vulnerabilities, such as CWE-264 (Permissions, Privileges, and Access Controls) and CWE-476 (Null Pointer Dereference). Detecting these issues often requires complex dependency analysis or additional external information. For instance, identifying CWE-264 vulnerabilities may necessitate an understanding of the software’s permission control scheme to evaluate whether the permissions used in the code are appropriate. A null pointer dereference vulnerability also requires cross-procedure information, which is not available within a single function, to make an accurate prediction.

\begin{tcolorbox}[breakable,width=\linewidth,colback=cyan!10, colframe=white,boxrule=0pt, left=1mm, right=1mm, top=1mm, bottom=1mm]
\noindent\textbf{Answer-3}: Leveraging the fundamental capabilities of code understanding acquired from pre-training on large-scale code corpora, PLMs generally perform well on vulnerabilities with obvious vulnerability patterns. However, they struggle with vulnerabilities that involve complex data and control dependencies, or those that require contextual information to be accurately identified. Notably, for certain complex vulnerabilities, PDBERT outperforms other models, suggesting that incorporating program dependencies during pre-training can better guide models in understanding code semantics. This suggests that future research could explore designing pre-training objectives tailored to specific code structures or vulnerability characteristics to further improve PLM performance in VD.
\end{tcolorbox}

\subsection{RQ4: How do PLMs perform under different code normalization rules?}\label{RQ4}

\begin{figure}[h]\centering
\includegraphics[width=0.45\textwidth]{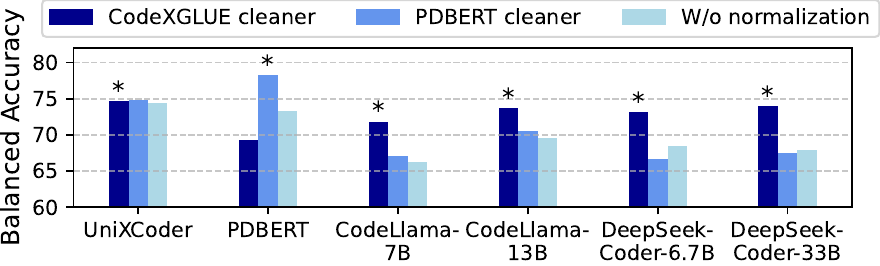}
\caption{PLMs' performance under different code normalization rules (* marks model's default normalization)} \label{fig:normalization}
\Description{Empty}
\end{figure}

Whitespace and newline characters in C/C++ are often used solely for formatting, which improves readability while preserving semantics. Benchmarks like CodeXGLUE~\cite{codexglue} often apply code normalization during preprocessing. However, different PLMs adopt varying normalization strategies during pre-training or fine-tuning. For example, CodeBERT removes all whitespace, \texttt{\small{\textbackslash t}}, and \texttt{\small{\textbackslash n}} characters, whereas PDBERT retains \texttt{\small{\textbackslash n}}. These differences raise the question of whether inconsistencies in input formatting impact model performance. In other words, can PLMs demonstrate robustness when there are differences in code normalization between training and testing? To investigate this, we compare model performance under three common normalization rules: (1) CodeXGLUE cleaner: removing all multiple whitespaces, \texttt{\small{\textbackslash t}}, and \texttt{\small{\textbackslash n}}; (2) PDBERT cleaner: removing multiple whitespaces and \texttt{\small{\textbackslash t}} but retains \texttt{\small{\textbackslash n}}; and (3) No normalization: using the original code. Each PLM is fine-tuned with its default normalization, and evaluated on the reconstructed test set under the above three normalization settings.

As shown in Fig.~\ref{fig:normalization}, PLMs generally perform best when the same normalization rules are applied during both training and testing, as they have learned to recognize patterns that closely resemble those in the test set. However, even minor perturbations in input formatting can cause significant performance degradation. For example, when evaluating PDBERT, removing \texttt{\textbackslash n} results in a performance drop of approximately 6.9\%, highlighting the model’s sensitivity and lack of robustness to subtle variations in code formatting.

\begin{tcolorbox}[breakable,width=\linewidth,colback=cyan!10, colframe=white,boxrule=0pt, left=1mm, right=1mm, top=1mm, bottom=1mm]
\noindent\textbf{Answer-4}: Inconsistency in code normalization during training and testing leads to significant performance degradation of PLMs, indicating that they lack robustness to minor perturbations in code format.
\end{tcolorbox}

\subsection{RQ5: Are PLMs capable of understanding abstracted vulnerability code?
}\label{sec:texual}

\begin{figure}[h]\centering
\includegraphics[width=0.45\textwidth]{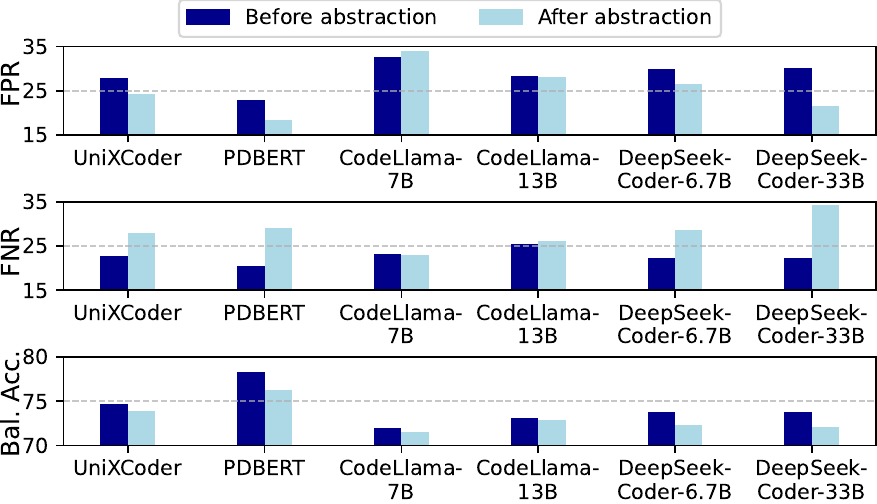}
\caption{PLMs' performance before and after abstraction} \label{fig:abstraction}
\Description{Empty}
\end{figure}

For privacy concerns, developers may mask the name of identifiers in their code when using PLMs for code analysis tasks. To examine whether PLMs can still understand vulnerable code without relying on textual information, we abstract all identifiers and strings in the code within the reconstructed test set. Specifically, we use Joern~\cite{Joern} to identify all variables, parameters, and strings within the functions, and replace them with abstracted forms such as \texttt{\small{VAR0}}, \texttt{\small{PARAM0}}, and \texttt{\small{STRING0}}, where the trailing numbers distinguish different entities. We then input both the original and abstracted functions into PLMs and compared their prediction results.

As shown in Fig.~\ref{fig:abstraction}, the false positive rate (FPR) of PLMs decreases after abstraction, indicating that textual information can sometimes mislead their predictions. On the other hand, the false negative rate (FNR) increases, suggesting that with reduced reliance on keywords, PLMs have lower confidence in detecting vulnerabilities, as key function or variable names may help them understand code semantics. Overall, the average balanced accuracy of PLMs drops by 1\%, indicating that while textual information does have some influence, its impact on PLM performance is relatively limited.
\begin{tcolorbox}[breakable,width=\linewidth,colback=cyan!10, colframe=white,boxrule=0pt, left=1mm, right=1mm, top=1mm, bottom=1mm]
\noindent\textbf{Answer-5}: PLMs are to some extent robust to code abstraction, which suggests that their vulnerability predictions do not rely solely on textual content. Instead, the fundamental code understanding capabilities learned from large code corpora, or the ability to analyze program dependency relationships acquired during pre-training, help them comprehend code logic and structural information for VD.
\end{tcolorbox}
\subsection{RQ6: Are PLMs capable of detecting vulnerable code after semantic-preserving transformation?}\label{RQ6}

\begin{figure}[h]\centering
\includegraphics[width=0.45\textwidth]{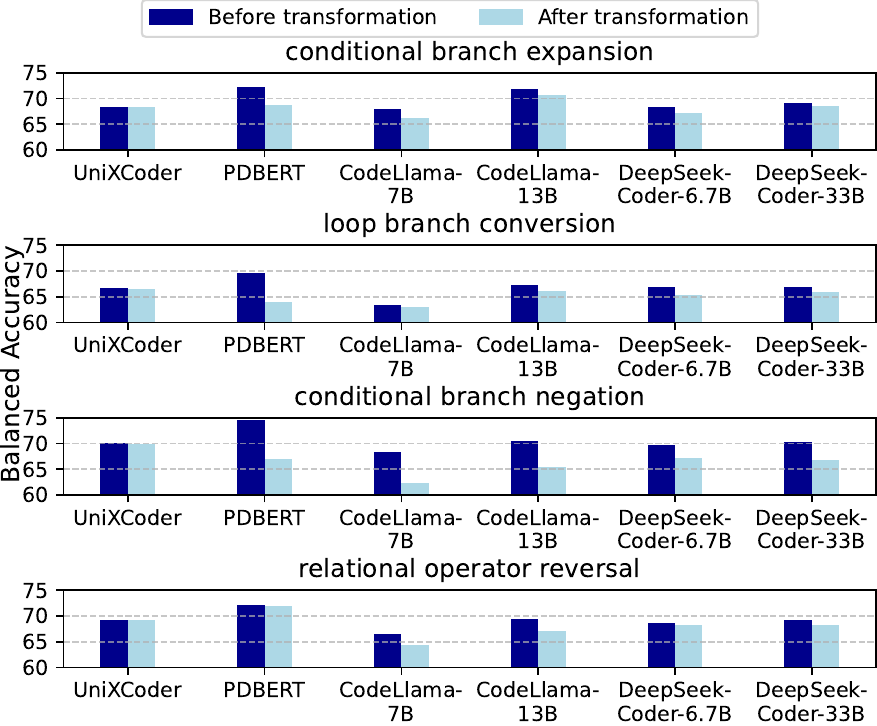}
\caption{PLMs' performance before and after semantic-preserving transformations} 
\label{fig:semantic_preserving}
\Description{Empty}
\end{figure}

Effective PLMs should demonstrate robustness against semantic-preserving transformations for two key reasons. First, the widespread adoption of OSS has amplified the risk of software vulnerabilities, as insecure code is often cloned and reused with minor syntactic modifications during copy-paste by less experienced developers~\cite{VGRAPH, cpminer}. Second, attackers may launch adversarial example attacks by deliberately applying semantic-preserving transformations to modify vulnerable code in ways that evade detection~\cite{yang2022natural,zhang2023challenging}. 
However, detecting such vulnerabilities is challenging: although the underlying semantic flaws persist, the code syntax are altered. 

\begin{figure}[h]
    \centering
    \subcaptionbox{Conditional branch negation\label{fig:rule1}}{
        \begin{minipage}[b]{0.45\textwidth}
            \includegraphics[width=1\textwidth]{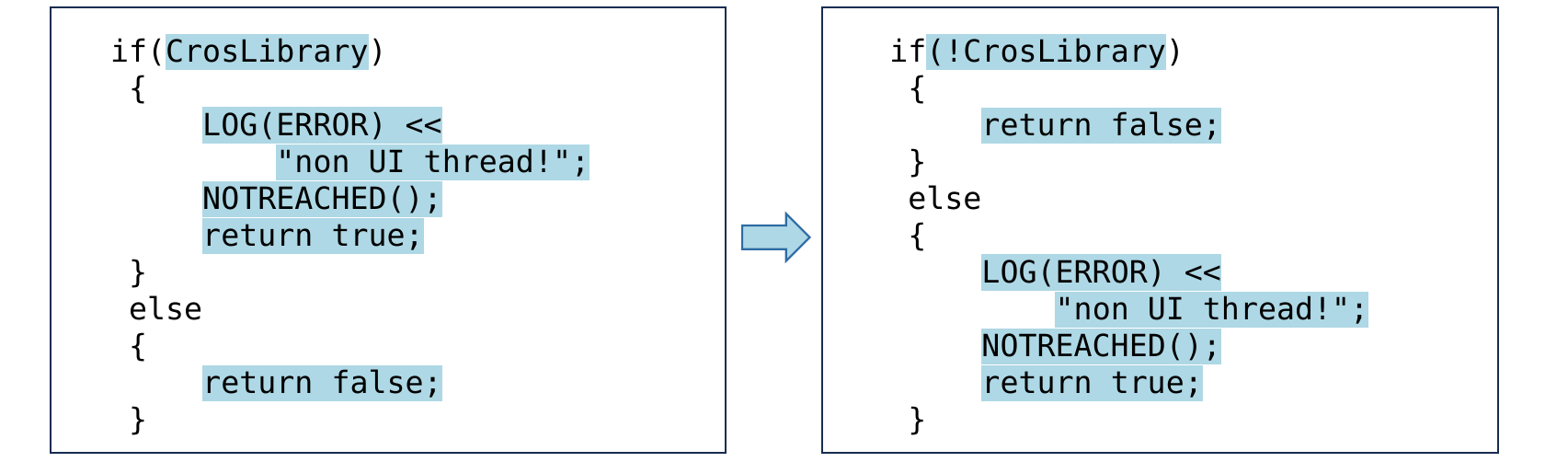}
        \end{minipage}
    }
    \subcaptionbox{Relational operator reversal\label{fig:rule2}}{
        \begin{minipage}[b]{0.45\textwidth}
            \includegraphics[width=1\textwidth]{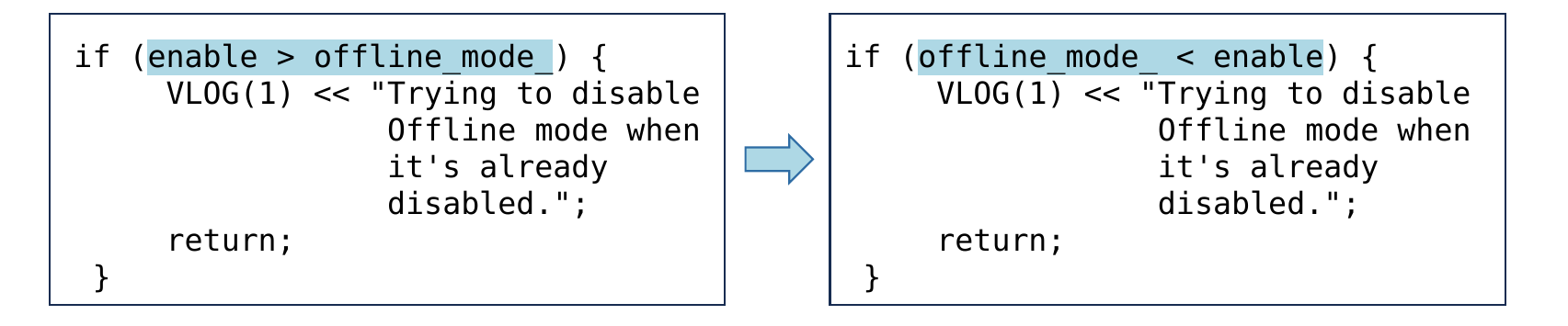}
        \end{minipage}
    }
    \subcaptionbox{Conditional branch expansion\label{fig:rule3}}{
        \begin{minipage}[b]{0.45\textwidth}
            \includegraphics[width=1\textwidth]{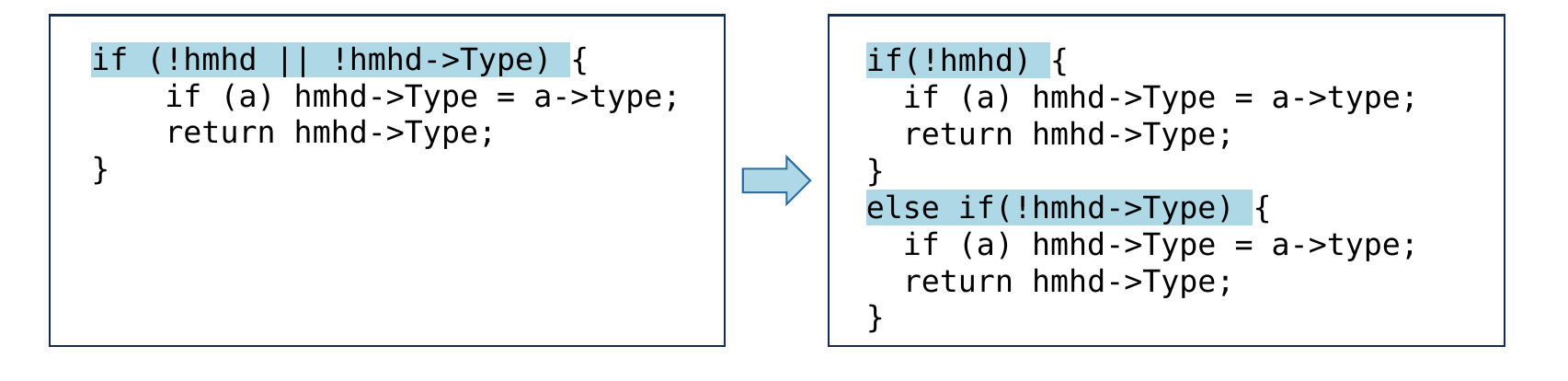}
        \end{minipage}
    }
    \subcaptionbox{Loop branch conversion\label{fig:rule4}}{
        \begin{minipage}[b]{0.45\textwidth}
            \includegraphics[width=1\textwidth]{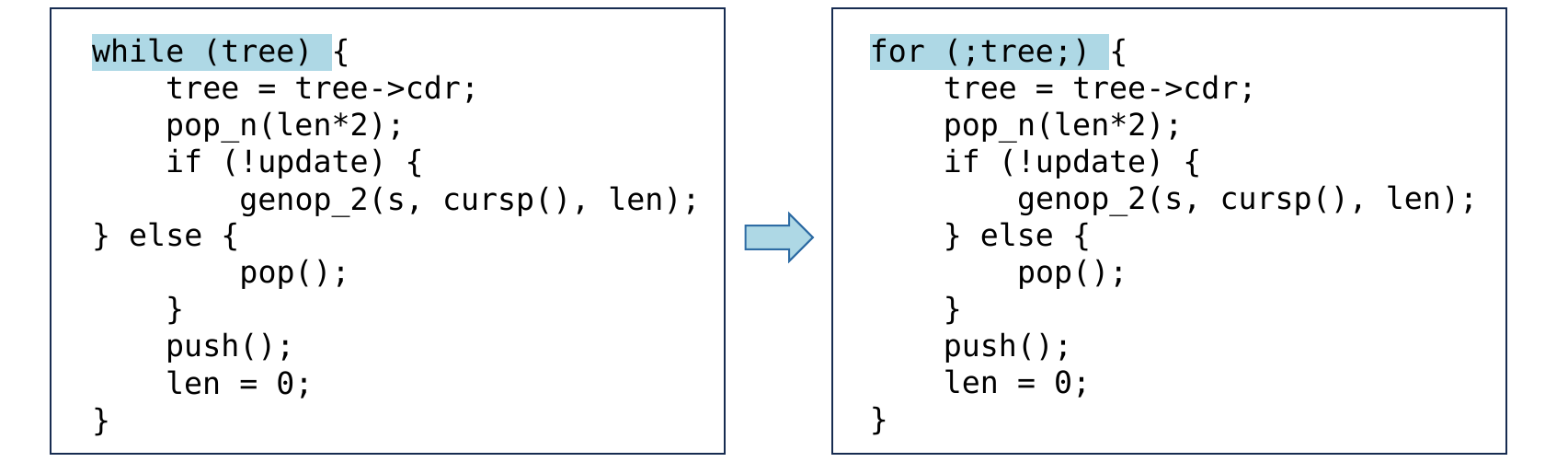}
        \end{minipage}
    }
    \caption{Semantic-preserving transformations}
    \vspace{-0.25in}
    \label{fig:trans_rule}
    \Description{Empty}
\end{figure}

To evaluate whether PLMs can produce consistent predictions for vulnerable code clone/reuse
, as shown in Fig.~\ref{fig:trans_rule}, we apply four types of semantic-preserving transformations (i.e., conditional branch negation, conditional branch expansion, loop branch conversion, and relational operator reversal) to all functions in the reconstructed test set. Specifically, if a transformation rule matches $M$ positions in a function, we apply the rule to each position individually, generating $M$ transformed functions. In total, based on 23,144 original functions, we generate 48,182 semantically equivalent functions. We evaluate PLMs on both the original and transformed test sets and report their performance for each type of transformation.

As shown in Fig.~\ref{fig:semantic_preserving}, the performance of PLMs degrades to varying degrees on different types of transformed data, whereas UniXCoder demonstrates strong robustness to code transformations with an average performance drop of only 0.1\% after transformation. A reasonable explanation is that its pre-training phase incorporates a multi-modal contrastive learning task, which helps it maintain consistent predictions for perturbed positive samples. In particular, PLMs show the highest robustness to the conditional branch expansion transformation, with an average performance decrease of only 1.3\% after transformation. A possible reason is that breaking down compound conditions into separate branches provides more detailed logical steps while preserving the original semantic structure. However, PLMs exhibit the lowest robustness to the conditional branch negation transformation, with an average performance decrease of 15.8\%. This may be because this transformation significantly changes the positions of code snippets, which could cause easily trackable data flows to be separated by unrelated code, making it difficult for PLMs to reliably extract long-distance dependencies. Additionally, PLMs have moderate robustness to relational operator reversal and loop branch conversion transformations. While reversed comparisons do not change the code logic, they might still pose challenges in model parsing due to variations in tokenization or the interpretation of relational operators. Similarly, although the syntactic differences between \texttt{\small{while}} and \texttt{\small{for}} loops are relatively small, initializing, condition-checking, and incrementing within a single line (as in \texttt{\small{for}} loops) versus being spread out (as in \texttt{\small{while}} loops) may introduce complexity in semantic parsing. 
\begin{tcolorbox}[breakable,width=\linewidth,colback=cyan!10, colframe=white,boxrule=0pt, left=1mm, right=1mm, top=1mm, bottom=1mm]
\noindent\textbf{Answer-6}: 
Most PLMs exhibit varing degrees of performance drop in semantic-preserving code, indicating they are not yet reliable for detecting potential vulnerable code clones and reuse in real-world scenarios. In contrast, UniXCoder mitigates the impact of transformed samples and maintains consistent performance by introducing multi-modal contrastive learning during pre-training. This highlights the potential for future VD research to design pre-training tasks that enhance adversarial robustness, thereby improving resilience to code transformations.
\end{tcolorbox}
\subsection{RQ7: To what extent do implicit labeling errors caused by token truncation affect the performance of Code SLMs?}\label{RQ7}

\begin{figure}[h]\centering
\includegraphics[width=0.45\textwidth]{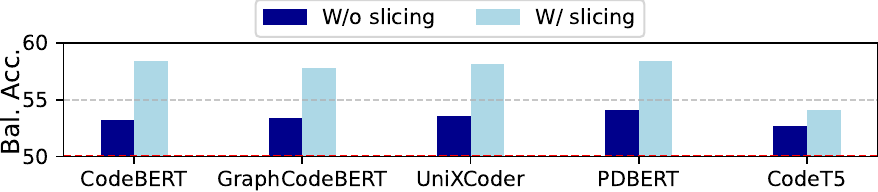}
\caption{Code SLMs' performance before and after slicing} \label{fig:slicing}
\vspace{-0.1in}
\Description{Empty}
\end{figure}

In addition to labeling errors during data collection, we identify an implicit labeling error that can result in unintentional data poisoning during token truncation. Due to the limited context window size (e.g., 512 tokens) of most Code SLMs, input functions are often truncated when fed into the models. This truncation prevents the models from accessing complete code context, forcing them to make predictions based on incomplete or less vulnerability-relevant code segments. Consequently, model training can be misled by these noisy labels, potentially compromising the training process.

\begin{figure}[h]\centering
\includegraphics[width=0.45\textwidth]{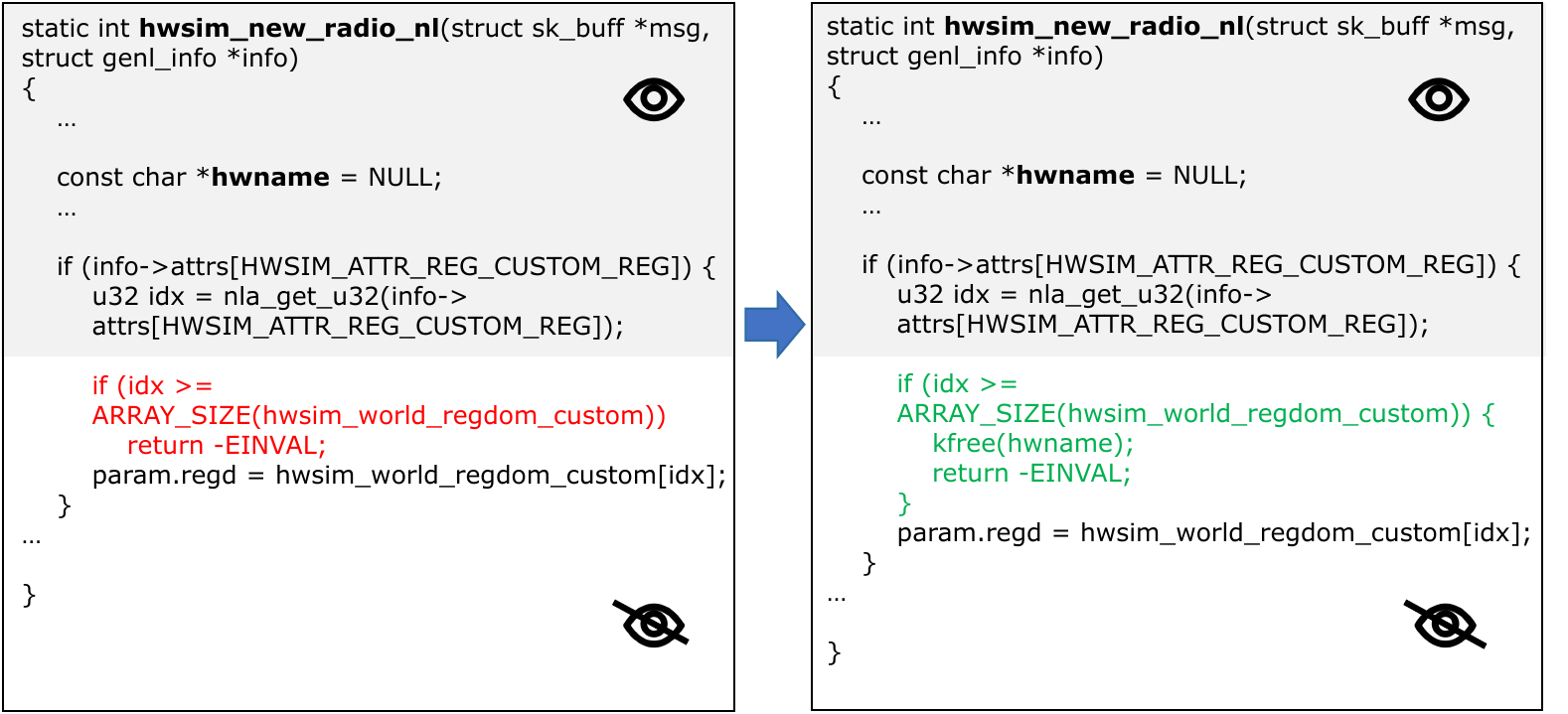}
\caption{Limited visible part caused by truncation} \label{fig:label_error}
\vspace{-0.05in}
\Description{Empty}
\end{figure}

For example, Fig.~\ref{fig:label_error} shows a memory leakage vulnerability-patched commit (ID 0ddcff4 in Linux Kernel~\cite{Linux_commit})
. Due to truncation, the visible input portions of the vulnerable and non-vulnerable function appear identical to the Code SLMs. However, their ground-truth labels are opposite, meaning that the same inputs correspond to opposite labels. This will confuse model training, resulting in unsatisfactory model performance. By analyzing 5,480 patch pairs in PrimeVul, we find that 1,473 pairs (27\%) have this issue. Although one naive way to avoid the labeling error caused by truncation is to simply exclude functions exceeding the maximum input size of Code SLMs, this limits the size of training data, leading to unsatisfactory model performance and making the learned models hard to deploy in real-world systems. 

Since not all statements in vulnerable functions are related to vulnerabilities, we perform code slicing to reduce the input length while preserving the core parts related to vulnerabilities as much as possible, thus minimizing the occurrence of labeling errors and enabling the model to learn more complete vulnerability patterns in functions. Specifically, we use tree-climber~\cite{treeclimber} to extracting both data and control flow of each patch pair in PrimeVul. To identify the core lines related to vulnerabilities, inspired by SySeVR~\cite{SySeVR}, we begin with designating lines that include API function calls, array usage, pointer usage, and arithmetic expressions as anchor lines. Then, to ensure all relevant dependencies are extracted, we perform slicing bidirectionally by forwarding and backtracking from the anchored lines. Finally, to prevent the total token count for sliced code lines from exceeding the maximum input size of the Code SLMs (i.e., 512), we implement a length checker that stops slicing when the token limit is reached. 

After slicing, we fine-tune the Code SLMs on both original paired training set of PrimeVul and sliced training set, and evaluate the performance of Code SLMs on their corresponding paired test set. As shown in Fig.~\ref{fig:slicing}, Code SLMs fine-tuned on the sliced training set achieve an average performance improvement of 4\% compared to those trained on the original paired training set. This supports our earlier hypothesis that token truncation introduces a substantial amount of poisoned data during training, ultimately degrading model performance. It is important to note that although SOTA LLMs have a larger context window size to support function-level VD, the issue of labeling errors caused by token truncation still remains in repo-level VD tasks, which requires an understanding of the entire codebase in real-world projects. 
\begin{tcolorbox}[breakable,width=\linewidth,colback=cyan!10, colframe=white,boxrule=0pt, left=1mm, right=1mm, top=1mm, bottom=1mm]
\noindent\textbf{Answer-7}: The limited context window size of Code SLMs not only hinders their ability to capture complete information from the code but also unintentionally introduces labeling errors during truncation, which disrupts model training. Code slicing, which retains the most vulnerability-related lines, effectively reduces input length and enables Code SLMs to focus more effectively on learning vulnerability patterns.
\end{tcolorbox}

\subsection{RQ8: How do PLMs perform under different prompt types and settings?}\label{prompt_ST}

\begin{figure}[h]\centering
\includegraphics[width=0.45\textwidth]{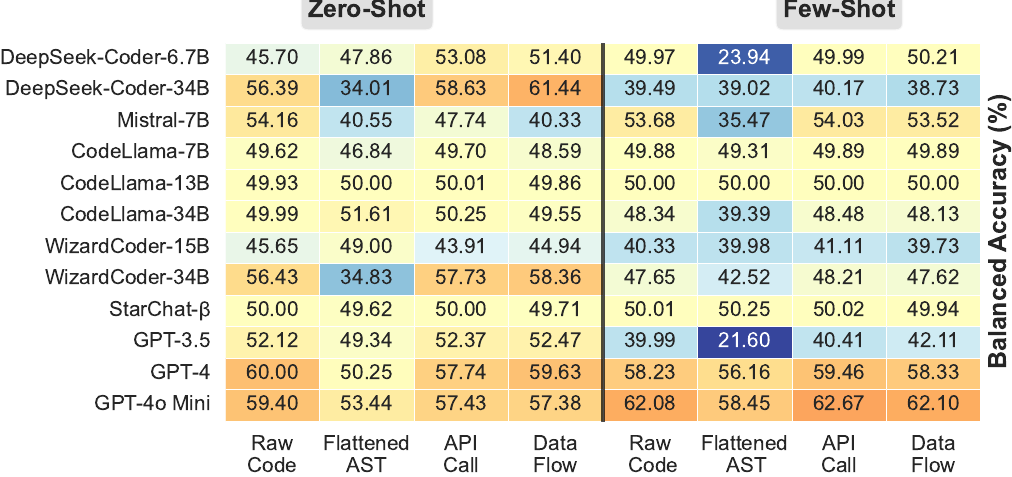}
\caption{PLM performance by prompt settings and types} 
\label{fig:llm_performance_heatmap}
\Description{Empty}
\end{figure}

In addition to the best performance of LLMs across all prompt settings and types reported in our prompting-based evaluations (Section~\ref{RQ1}), 
this section provides a detailed analysis of how different prompt configurations influence model performance. As shown in Fig.~\ref{fig:llm_performance_heatmap}, with respect to prompt types, GPT-4o Mini achieves its highest performance when API calls and few-shot examples are incorporated into the prompts. The effectiveness of adding supplementary structural information to prompts depends heavily on the model’s inherent code understanding capabilities. For instance, models like DeepSeek-Coder typically benefit from the inclusion of API call and data flow, as such additions help reinforce code logic and clarify dependencies. This is likely because DeepSeek-Coder already possesses a solid understanding of raw code. On the other hand, Code LLMs such as CodeLlama and StarChat-$\beta$, which perform at or below random guessing on raw code, often struggle when confronted with these added logic chains, making code comprehension even more challenging.

Regarding prompt settings, few-shot in-context learning enhances the performance of LLMs (e.g., GPT-4o Mini) by leveraging extended context windows and reference examples. This approach serves as implicit guidance, encouraging models to follow instructions more effectively~\cite{ICT}. However, for Code LLMs such as DeepSeek-Coder and WizardCoder, which already show strong performance in zero-shot scenarios, introducing a limited number of few-shot examples may disrupt decision-making~\cite{ACL}. The complexity and variability of vulnerabilities can make a small set of few-shot examples insufficient to convey the necessary information. Even worse, it may lead the model to overfit on simplistic patterns from provided examples, rather than learning generalizable features.

\begin{tcolorbox}[breakable,width=\linewidth,colback=cyan!10, colframe=white,boxrule=0pt, left=1mm, right=1mm, top=1mm, bottom=1mm]
\noindent\textbf{Answer-8}: The effectiveness of different prompt settings 
and types 
depends on a model’s inherent code understanding. Models like GPT-4o Mini benefit from added structural information when they already exhibit a fundamental understanding of raw code, while more complex prompts may hinder weaker models. Given the complex and diverse vulnerability patterns that are difficult to capture with limited few-shot examples, in-context learning may mislead the model to overfit superficial patterns.
\end{tcolorbox}

\section{Discussion}
This paper differentiates itself from previous work by revisiting the true capabilities of PLMs for VD, addressing the shortcomings of existing VD evaluations. We argue that the conclusions drawn from previous, biased evaluations are meaningless, as they lead to a significant gap between reported performance and practical effectiveness. This is a key point emphasized throughout our work.

For dataset selection, to minimize data leakage, we limit our dataset to commits dated after the knowledge cutoff dates of all evaluated LLMs. While some undetectable forms of leakage may still exist, addressing them is beyond the scope of this study.

For model selection, we acknowledge that some open-source reasoning models (e.g., Qwen3~\cite{qwen3}) have shown strong performance in the software engineering tasks. However, their very recent knowledge cutoff date conflicts with the scarcity of latest vulnerabilities, making it difficult to collect enough recent data to ensure the accuracy and comprehensiveness of evaluation results. We also clarify that while resource and budget constraints prevent us from evaluating all SOTA LLMs, our evaluation framework can help future research evaluate stronger LLMs on more recent VD data.

Although our findings conclude that Code SLMs (e.g., PDBERT), which are trained on specialized code pretraining tasks, generally outperform LLMs trained on general language modeling, their applicability is constrained to function-level VD due to the limited context window size. In contrast, LLMs provide a larger context window, but their ability to perform longer function-level or cross-procedure VD still needs significant improvement.
\section{Conclusion}
This work presented an extensive evaluation of the vulnerability detection capabilities of representative PLMs. The findings revealed that PLMs incorporating pre-training tasks designed to capture the syntactic and semantic patterns of code outperform those solely pre-trained on general language modeling tasks. However, they remain sensitive to code normalization and transformation, and face challenges in handling vulnerabilities with complex dependencies or those exceeding the context window size.

\bibliographystyle{ACM-Reference-Format} 
\bibliography{RevisitVD.bib}    

\appendix

\section{Ethical Considerations}

Our work evaluates existing pre-trained language models on the vulnerability detection task. All experiments are conducted using publicly available data, models, and source code. Since no new vulnerabilities are generated or identified in this work, no ethical issues will arise.

\section{Models}\label{sec:models}

\subsection{Small Language Models for Code (Code SLMs)}

\noindent\textbf{CodeBERT}~\cite{codebert} is a bimodal pre-trained encoder-based Transformer model that utilizes a multi-layer bidirectional Transformer-based architecture to learn general-purpose representations for programming language (PL) and natural language, achieving good performance in the natural language code search and code documentation generation through incorporating a replaced token detection objective.

\noindent\textbf{GraphCodeBERT}~\cite{graphcodebert} is a pre-trained encoder-based Transformer model that incorporates inherent semantic structure of code using data flow rather than traditional syntactic structures like ASTs to enhance code understanding. By employing a multi-layer bidirectional Transformer-based architecture and introducing two structure-aware pre-training tasks (i.e., edge prediction and node alignment), it achieves satisfactory performance on tasks such as code search, clone detection, code translation, and code refinement.

\noindent\textbf{UniXCoder}~\cite{UniXcoder} is a unified cross-modal pre-trained model designed for PL, which optimizes code understanding and generation tasks by utilizing mask attention matrices and prefix adapters, as well as integrating cross-modal contents such as AST and code comments to achieve code fragment representation learning. It demonstrates advanced performance across five code-related tasks and a new zero-shot code-to-code search task, highlighting the benefits of leveraging both comments and AST structures in its architecture.

\noindent\textbf{PDBERT}~\cite{pdbert} is a pre-trained encoder-based Transformer model designed to enhance dependency analysis in software by introducing two novel pre-training objectives: statement-level control dependency prediction and token-level data dependency prediction. PDBERT significantly improves understanding of vulnerable code and achieves strong performance on vulnerability detection, classification, and assessment tasks as well as program dependence analysis.

\noindent\textbf{CodeT5}~\cite{codet5} is a unified pre-trained encoder-decoder Transformer model designed for effectively leveraging code semantics from developer-assigned identifiers. It introduces an identifier-aware pre-training task to differentiate and recover identifiers and utilizes user-written code comments in a bimodal dual generation task for better alignment between nature language and PL, resulting in significant performance improvements over previous methods in code understanding tasks (e.g., code defect detection and clone detection) and various generation tasks.

\subsection{Large Language Models for Code (Code LLMs)}

\noindent\textbf{CodeLlama}~\cite{codellama} is initialized from LLaMA 2 and featured by a multitask objective combining autoregressive and causal infilling prediction, enabling various tasks such as code completion and docstring generation. It captures long input contexts by adjusting the parameters of RoPE positional embeddings. Further fine-tuned on a blend of proprietary instruction data and a machine-generated self-instruct dataset, the helpful and safe CodeLlama-Instruct models better follow human instructions and achieve excellent performance among open models on several code benchmarks, including HumanEval and MBPP.

\noindent\textbf{DeepSeek-Coder}~\cite{deepseekcoder} is initialized from the DeepSeek LLM. By pre-training with a fill-in-the-middle approach on a combination of project-level source code and natural language corpora in English and Chinese, DeepSeek-Coder models enhance their versatility and applicability across various coding scenarios. Extending the input sequence length by reconfiguring RoPE parameters further strengthens the models’ reliable ability to handle long-context tasks, such as cross-file code completion. Incorporating high-quality, helpful, and impartial human instruction data, DeepSeek-Coder-Instruct significantly narrows the performance gap between OpenAI’s GPT-4 and open-source models on a wide range of code-related benchmarks.

\noindent\textbf{StarChat-$\beta$}~\cite{starchat} is initialized from StarCoderBase, which was pre-trained on the Stack dataset covering over 80 programming languages. Fine-tuning StarCoderBase on 35B Python tokens enables StarCoder to perform well in Python and other languages. To further enhance its capabilities as a coding assistant, StarChat-$\beta$ is fine-tuned from StarCoderPlus, which was trained on an uncensored variant of the OpenAssistant-Guanaco dataset, making the model more helpful for coding tasks.

\noindent\textbf{WizardCoder}~\cite{wizardcoder} is initialized from the StarCoder series. It applies the Evol-Instruct method, proposed by WizardLM, to construct the instruction-following training set for fine-tuning. Comprehensive experiments on four code generation benchmarks (i.e., HumanEval, HumanEval+, MBPP, and DS1000) reveal the model’s exceptional capabilities compared to other open-source code LLMs.

\end{document}